\documentclass[11pt]{article}
\pdfoutput=1
\usepackage{amsmath, amsfonts, amssymb}
\usepackage{comment}
\usepackage{graphicx}
\usepackage{psfrag}
\usepackage{amsthm}
\usepackage[usenames,dvipsnames,svgnames,table]{xcolor}
\usepackage{enumerate}
\usepackage{tcolorbox}
\usepackage{mathtools}
\usepackage{soul}
\usepackage{subfigure}
\usepackage{mathrsfs}  
\usepackage{a4wide}
\usepackage{booktabs}
\usepackage{tikz}
\usepackage{tikz-cd}
\usepackage{makecell}
\usepackage{multicol}
\usepackage{braket}

\usepackage{color}
\definecolor{dark-gray}{gray}{0.20}
\definecolor{gray}{gray}{0.30}
\definecolor{light-gray}{gray}{0.80}
\definecolor{dark-red}{rgb}{0.7,0,0}
\definecolor{dark-green}{rgb}{0.1,0.4,0}
\definecolor{dark-blue}{rgb}{0.3,0.3,0.7}
\definecolor{light-blue}{rgb}{0.8,0.8,1}
\definecolor{swamp}{RGB}{240, 199, 197}

\usepackage{pifont}

\usepackage{setspace}

\newcommand{\be}{\begin{equation}}
\newcommand{\ee}{\end{equation}}

\def\be{\begin{equation}}
\def\ee{\end{equation}}
\def\bea{\begin{eqnarray}}
\def\eea{\end{eqnarray}}

\newcommand{\arXivold}[2]{\href{https://arxiv.org/abs/#1}{\tt arXiv:#1 [#2]}}

\def\simleq{\; \raise0.3ex\hbox{$<$\kern-0.75em
		\raise-1.1ex\hbox{$\sim$}}\; }
\def\simgeq{\; \raise0.3ex\hbox{$>$\kern-0.75em
		\raise-1.1ex\hbox{$\sim$}}\; }

\numberwithin{equation}{section}

\usepackage{jheppub} %
\usepackage{orcidlink}
\hypersetup{
	colorlinks=true,
	linkcolor=dark-blue,
	citecolor=dark-red,
	urlcolor=dark-green
}

\title{On Particle Production from Bubble Collisions in the Non-runaway Regime}

\author[a]{Fayez Abu-Ajamieh\,\orcidlink{0000-0002-2814-0190},}
\author[b]{Amine Ahriche\,\orcidlink{0000-0003-0230-1774},}
\author[c]{Sagar Modak\,\orcidlink{0009-0009-1057-9283}}
\author[d]{and Nobuchika Okada\,\orcidlink{0000-0002-3798-773X}}

\affiliation[a]{Formerly: Indian Institute of Science, Bangalore, India}
\affiliation[b]{Department of Applied Physics and Astronomy, University of Sharjah,\\ P.O. Box 27272, Sharjah, UAE}
\affiliation[c]{Centre for High Energy Physics, Indian Institute of Science,\\ C V Raman Avenue, Bengaluru 560012, India}
\affiliation[d]{Department of Physics and Astronomy, University of Alabama,\\ Tuscaloosa, AL 35487, USA}

\emailAdd{fayezabuajamieh@gmail.com}
\emailAdd{ahriche@sharjah.ac.ae}
\emailAdd{sagarmodak@iisc.ac.in}
\emailAdd{okadan@ua.edu}

\abstract{Particle production from bubble collisions during First-Order Phase Transitions (FOPT) has been studied as an efficient mechanism for baryogenesis and dark matter production, but
 existing semi-analytical treatments assume runaway bubbles with wall velocity $v_w\to 1$. 
 We generalize this treatment to arbitrary $v_w< 1$, relevant for scenarios such as 
 electroweak baryogenesis and transitions with significant wall friction. We first extend 
 the analytical trapping-equation approach to finite-thickness boosted kink profiles for 
 both elastic and inelastic collisions, then go beyond this by numerically solving the full
 Klein-Gordon equation without assuming $SO(1,1)$ symmetry or thin walls, across a broad grid of $v_w$ and potential shapes. We find the analytical generalization underestimates the
 true efficiency factor by orders of magnitude away from $v_w\to 1$, and provide new semi-analytic fit functions consisting of a power-law continuum plus a resonance peak, which reproduce the main spectral features of the numerical results across the parameter range studied. Applying these to a toy dark 
 sector model, we show particle production is strongly suppressed as $v_w$ decreases from 
 unity, enhanced for lighter produced particles, and that the relative contribution from 
 the oscillatory peak differs from prior runaway-regime estimates. We also discuss limitations
 of the trapping equation and the off-shell effective-action truncation underlying this formalism.}

\begin{document}

\maketitle

\section{Introduction}\label{intro}

The topic of First-Order Phase Transitions (FOPTs) has a long history~\cite{Coleman:1977py, Callan:1977pt, Linde:1980tt, Hawking:1982ga, Turner:1992tz, Abu-Ajamieh:2026ndt, Abu-Ajamieh:2025sul, Abu-Ajamieh:2024xic, Abu-Ajamieh:2024gaw}. In a FOPT, bubbles of true vacuum nucleate, expand and collide in the sea of false vacuum, and although in the Standard Model (SM) the transition is crossover, BSM extensions with low-energy Effective Field Theories (EFTs)~\cite{Abu-Ajamieh:2020yqi, Abu-Ajamieh:2021vnh, Abu-Ajamieh:2021egq, Abu-Ajamieh:2022ppp} can accomplish a strong FOPT, such as including a scalar singlet extensions to the SM~\cite{Choi:1993cv, Ham:2004zs,Espinosa:1993bs, Ahriche:2007jp, Barger:2008jx,  Kurup:2017dzf, Ahriche:2014jna}, the Two-Higgs Doublet Model (2HDM)~\cite{Turok:1990in, McLerran:1990zh,Jain:1993an, Cline:1996mga, Fromme:2006cm, Cline:2011mm, Basler:2016obg, Aoki:2021oez, Abu-Ajamieh:2025zcv}, non-local QFTs~\cite{Ghoshal:2022mnj, Abu-Ajamieh:2023roj, Abu-Ajamieh:2023txh, Abu-Ajamieh:2024woy, Abu-Ajamieh:2024egb}, extra dimensions~\cite{Creminelli:2001th, Abu-Ajamieh:2016ulh, Bunk:2017fic, Abu-Ajamieh:2017khi, Megias:2018sxv, Abu-Ajamieh:2018brk}, and supersymmetry~\cite{Carena:1997ki,Carena:1997gx, Cline:1998hy, Espinosa:1996qw, Laine:2000kv}.

FOPTs have attracted significant attention over the years for several reasons. First, a FOPT can cause a significant departure from thermal equilibrium, which is one of the three Sakharov conditions required to achieve baryogenesis~\cite{Sakharov:1967dj}. Second, a FOPT can lead to the production of potentially detectable stochastic Gravitational Waves (GWs)~\cite{Hogan:1983ixn, Witten:1984rs, Hogan:1986dsh, Kosowsky:1991ua, Kosowsky:1992rz, Kosowsky:1992vn, Kamionkowski:1993fg}. Third, a FOPT can generate primordial magnetic fields~\cite{Vachaspati:1991nm, Cheng:1994yr, Kibble:1995aa, Sigl:1996dm, Ahonen:1997wh}. 

The dynamics of bubble expansion itself is governed by the balance between the pressure driving the walls outward and the  friction they experience from the surrounding plasma~\cite{Bodeker:2009qy, Bodeker:2017cim}. For cases in which the friction  is negligible or absent, such as in strongly supercooled transitions, in transitions proceeding via tunneling in a cold  sector, or in the absence of gauge-boson splitting radiation, the walls fail to reach a terminal velocity and instead ``run away", retaining an ever-growing fraction of the released vacuum energy as kinetic and gradient energy, with the boost 
factor $\gamma_w$ growing linearly with the bubble radius. It is in precisely this runaway regime that bubble collisions  become most consequential: the kinetic term of the scalar field's equation of motion dominates over the potential-gradient  term during the collision, so that the colliding wall profiles approximately superpose rather than dissipate. This free-passage behavior underlies the so-called ``trapping equation"~\cite{Jinno:2019bxw}, which reduces the dynamics of the post-collision field to a single light-front variable $s=\sqrt{t^2-x^2}$ and determines whether the field subsequently bounces back into the false vacuum (an elastic collision) or settles into the true vacuum (an inelastic collision). This same  runaway behavior is also central to gravitational-wave phenomenology: when the bubble walls retain the dominant share of the  transition's energy, the GW signal is sourced directly by the collision of the scalar field walls~\cite{Kosowsky:1992rz,Kosowsky:1992vn,Kamionkowski:1993fg, Huber:2008hg,Jinno:2016vai}, whereas if the energy is instead transferred efficiently to the surrounding plasma, the dominant GW sources become sound waves~\cite{Hindmarsh:2013xza, Hindmarsh:2019phv} and turbulence, or, in the absence of strong plasma interactions, dynamic distributions of free-streaming particles~\cite{Jinno:2022mie}.

In addition, it was suggested in Ref.~\cite{Watkins:1991zt} that colliding bubble walls in a strong FOPT could produce heavy scalar particles in the form of ultrarelativistic collisions. This mechanism was first applied to baryogenesis in Ref.~\cite{Konstandin:2011ds}, while~\cite{Falkowski:2012fb} provides analytical expressions for the number of particles produced in such collisions for both elastic and inelastic cases, which were further used to design a mechanism for the production of heavy Dark Matter (DM) particles via the Higgs portal. The same mechanism of producing heavy DM particles in  bubble collisions turned out to be useful in other contexts, e.g., for nonthermal heavy DM production~\cite{Giudice:2024tcp} or leptogenesis~\cite{Cataldi:2024pgt}. Thus, the efficiency factor of such production turns out to be a general element of any FOPT model with strong enough self- or portal coupling. Refs.~\cite{Mansour:2023fwj,Shakya:2023kjf} used the trapping equation~\cite{Jinno:2019bxw} to obtain semi-analytical expressions for efficiency factors for the production of relativistic particles from bubble collisions by approximating the numerical solution in the runaway regime $v_{w} \rightarrow 1$, where $v_{w}$ is the bubble wall velocity. Furthermore, it was shown in Ref.~\cite{Inomata:2024rkt} that the particles produced in bubble collisions provide an additional contribution to the stochastic GW background, which is determined by the quadrupole anisotropy of their momentum distribution in the frame of reference associated with the cosmological background. This effect dominates over the conventional ``bulk flow'' and envelope approximation contributions at long wavelengths, which allows one to distinguish it from the scalar contributions predicted by the sound wave and MHD turbulence mechanisms. Note that this provides another channel for GW generation in addition to the direct collisions of bubble walls, sound waves, and MHD turbulence. 

A major limitation of~\cite{Mansour:2023fwj,Shakya:2023kjf}, and indeed essentially all of what has been discussed above, is that these treatments were limited to ultrarelativistic bubble walls $v_{w} \rightarrow 1$. While certainly a case of interest in the context of many FOPT models (particularly supercooled transitions), runaway bubbles are not the only case of interest, and the assumption of free passage that underlies the trapping equation itself is only strictly speaking justified in this limit. For example, subsonic deflagrations are traditionally favored for electroweak (EW) baryogenesis, although viable baryogenesis can also be possible at higher wall velocities~\cite{Kozaczuk:2015owa,Cline:2020jre}. For non-runaway 
collisions, sound waves are the dominant source of GWs compared with bubble collisions in runaway bubbles, changing the characteristic peaks and spectral shape of GWs~\cite{Hindmarsh:2013xza, Hindmarsh:2019phv}. In addition, in many scenarios of confinement-deconfinement transitions, a large number of degrees of freedom could become massive, providing significant friction leading the bubbles to reach a terminal velocity. More generally, any scenario where the bubble walls do not achieve pure runaway behavior (whether due to plasma friction, particle-production backreaction or the dynamics of the potential) falls outside the regime where existing efficiency-factor calculations apply. Therefore, a dedicated treatment of particle production from bubble collision in the regime where $v_{w} < 1$ is required, both as an extension of the particle-production literature and as an input into the broader program of characterizing the full multi-component GW signal, particle-physics implications, and cosmological consequences of realistic, non-idealized FOPTs. 

In this work, we attempt to generalize the treatment in Ref.~\cite{Mansour:2023fwj} to the non-runaway case. Our approach will be similar in that we primarily utilize numerical fitting, but we use less idealized assumptions. Specifically, we begin by generalizing the analytical treatment in Ref.~\cite{Falkowski:2012fb} to $v_{w} < 1$, then we proceed to numerically solve the full Klein-Gordon (KG) equation assuming the interaction starts at $t=0$. We do not assume any $SO(1,1)$ symmetry and treat 
the bubble wall thickness as finite. We compare our numerical results with the generalized analytical expressions and show that although they have acceptable accuracy in the ultrarelativistic regime, they nevertheless generalize poorly to the lower wall velocities. We derive generalized semi-analytic expressions for the efficiency factors for a broad range of $v_{w}$ and we compare our results with those in Ref.~\cite{Mansour:2023fwj}. We also investigate the impact of $v_{w}$ on the particle production rate and find that it is greatly enhanced as $v_{w} \rightarrow 1$. Along the way, we also revisit several simplifying assumptions retained even in the runaway-regime literature -- the step-function wall profile, the $SO(1,1)$ symmetry, and the validity of the trapping equation itself -- and find that relaxing them can have a non-negligible impact even in the ultrarelativistic limit, and we comment on the field reparametrization and gauge dependence inherent to the underlying off-shell formalism, an issue that affects this entire class of calculations independently of wall velocity.

This work is organized as follows. In Section~\ref{sec:Dynamics} we will review particle production dynamics. In Section~\ref{sec:AttemptedAnalytical} we will generalize the analytical expression presented in Ref.~\cite{Falkowski:2012fb} to $v_{w} < 1$. In Section~\ref{sec:ShakyaFit} we will review the results of~\cite{Mansour:2023fwj} for $v_{w} \rightarrow 1$ and in Section~\ref{sec:KGnumerical} we will generalize these results to the case of $v_{w} < 1$. In Section~\ref{sec:pp} we will investigate the impact of $v_{w}$ on the particle-production rate. In Section~\ref{sec:Issues} we will comment on the  issues associated with the usage of the off-shell prescription to describe particle production from bubble collisions and finally we will conclude in Section~\ref{sec:Discussion}.

\section{Review of particle production dynamics}
\label{sec:Dynamics}
We begin by reviewing the dynamics of particle production from bubble collisions that was used previously in the literature. Previous treatments assumed runaway bubbles with $v_{w} \rightarrow 1$. In this approximation, bubbles can be assumed to have infinitely thin walls, and the scalar field describing the walls is assumed to have an $SO(1,1)$ symmetry. This symmetry implies that the KG equation simplifies to an ordinary differential equation of one variable
\begin{equation}\label{eq:KG_SO}
\frac{d^{2}\phi}{ds^{2}} + \frac{1}{s}\frac{d\phi}{ds} +\frac{dV(\phi)}{d\phi} =0\,, 
\end{equation}
where $s = \sqrt{t^{2} - x^{2}}$ is the light-front coordinate. Bubble walls separate the true and false vacua, and when they collide, they superpose at the point of collision. If the collision is ultrarelativistic, the kinetic terms in the KG equation dominate over the potential-gradient term, and the field profiles approximately superpose. This free-passage approximation implies the trapping equation~\cite{Jinno:2019bxw}
\begin{equation}\label{eq:trapping}
\phi_{\text{after}} = 2\phi_{\rm T} - \phi_{\rm F}\,,
\end{equation}
where $\phi_{\rm T}$ ($\phi_{\rm F}$) is the field value in the true (false) vacuum. As pointed out in~\cite{Falkowski:2012fb, Mansour:2023fwj}, the field evolution after the collision is set by the details of the potential. In general, there are two limiting cases to consider: elastic and inelastic collisions. In elastic collisions, the two minima are almost degenerate and after the collision, the walls recede from one another re-establishing the false vacuum. This corresponds to the scalar climbing the potential barrier from the true vacuum back to the false one. Vacuum pressure will drive the walls to collide again several times until the scalar settles in the true vacuum. On the other hand, in inelastic collisions, which occur when the barrier separating the two minima is relatively shallow, the walls essentially stick together and the false vacuum is not reestablished.

Physically, the field configuration at the moment of collision can be understood as an ``excursion'' of the field to values beyond the true vacuum in the narrow region around the collision point, followed by a rapid oscillation back and forth across this excited value~\cite{Falkowski:2012fb}. Whether this oscillation is subsequently able to drive the field over the potential barrier back into the false vacuum (elastic case) or not (inelastic case) depends entirely on the shape of the potential in the vicinity of the barrier, and is precisely the question that the trapping equation, Eq.~(\ref{eq:trapping}), 
is designed to answer. It is worth stressing that the elastic scenario described here is itself an idealization: a fully elastic collision, as sketched above, is followed by repeated collisions of the receding and re-approaching walls, with each successive collision radiating away some fraction of the  wall energy into scalar waves and particle quanta, until the field eventually settles into the true vacuum~\cite{Falkowski:2012fb}. In this work, as in the analyses we build upon, we restrict our attention to a single collision event for both the elastic and inelastic cases, capturing the dominant physical effects while leaving a treatment of multiple sequential collisions to future work.

Before turning to the calculation of particle production itself, it is useful to note the amount of energy that is potentially available for conversion into particles. In the ultrarelativistic limit, the kinetic energy per unit area stored in a single approaching bubble wall of thickness $\ell_w$ boosted by a factor $\gamma_w$ is
\begin{equation}
\label{eq:Ew}
\frac{E_w}{A} = \frac{2}{3} v_{\phi}^{2}\,\frac{\gamma_w}{\ell_w}\,,
\end{equation}
which grows linearly with $\gamma_w$ and can therefore represent an enormous energy reservoir for sufficiently boosted runaway walls~\cite{Falkowski:2012fb, Watkins:1991zt}. This provides the basic energetic motivation for treating bubble collisions as a potentially efficient particle production mechanism: so long as the total energy radiated into any given particle species remains a small fraction of $E_w/A$, the calculation described below, which treats the produced particles as a perturbative source on top of the classical background field $\phi(z,t)$, neglecting the backreaction of particle production on the field's own evolution, remains self-consistent. If the naively computed particle yield for some channel instead approaches or exceeds $E_w/A$, this signals that backreaction can no longer be neglected and the results should be interpreted with due caution~\cite{Falkowski:2012fb}.

To describe the production of particles from bubble collisions, the colliding bubble walls and the subsequent oscillations are treated as a classical field $\phi(x,t)$, and the probability of its decay into particles is extracted from the imaginary part of the effective action. More specifically~\cite{Watkins:1991zt}
\begin{equation}\label{eq:pp_derivation1}
P_{\geq 1} = 1-\exp{(-2\text{Im}\Gamma[\phi])} \simeq 2 \text{Im}\Gamma[\phi]\,,
\end{equation}  
where $P_{\geq1}$ is the probability of producing at least one particle after the collision, and the action is given by
\begin{equation}\label{eq:pp_derivation2}
\Gamma[\phi] = \sum_{n=2}^{\infty}\frac{1}{n!}\int d^{4}x_{1}\cdots d^{4}x_{n} \Gamma^{(n)}(x_{1},\cdots,x_{n}) \phi(x_{1}) \cdots \phi(x_{n})\,,
\end{equation}
and $\Gamma^{(n)}(x_{1},\cdots,x_{n}) $ is the n-point 1PI. A textbook approximation is to truncate the action at the leading order, keeping only the 2-point function $\Gamma^{(2)}(x_{1},x_{2})$, and then Fourier-transform it to momentum space,
\begin{equation}\label{eq:pp_derivation3}
\text{Im}(\Gamma[\phi])  \approx \frac{1}{2}\int d^{4}x_{1} d^{4}x_{2} \phi(x_{1}) \phi(x_{2}) \int \frac{d^{4}p}{(2\pi)^{4}}e^{ip.(x_{1}-x_{2})}\text{Im}\Big( \tilde{\Gamma}^{(2)}(p^{2})\Big)\,.
\end{equation}
Inserting $\phi(x_i)=\int\frac{d^4k_i}{(2\pi)^4} e^{-ik_i\cdot x_i}\widetilde{\phi}(k_i)$ with $i=1,2$, and then performing the integrals, we find
\begin{equation}\label{eq:pp_derivation4}
\text{Im}(\Gamma[\phi])  \approx \frac{1}{2} \int \frac{d^{4}p}{(2\pi)^{4}}|\tilde{\phi}(p)|^{2}\text{Im}\Big( \tilde{\Gamma}^{(2)}(p^{2})\Big)\,.
\end{equation}
For plane bubbles moving in the $z$ direction, $\phi(x,t) = \phi(z,t)$ and the Fourier transform is $\tilde{\phi}(\vec{k},\omega) = (2\pi)^{2}\delta(k_{x})\delta(k_{y})\tilde{\phi}(k_{z},\omega)$. Substituting this in Eq.~(\ref{eq:pp_derivation4}) and performing the integral over $p$, then substituting back into Eq.~(\ref{eq:pp_derivation1}), and assuming that each excitation produces a pair of identical particles, we obtain the number of particles per unit area produced by bubble collisions
\begin{equation}\label{eq:pp_derivation5}
\frac{N}{A} \approx 2 \int\frac{dk d\omega}{(2\pi)^{2}}| \tilde{\phi}(k,\omega)|^{2}\text{Im}[\tilde{\Gamma}^{(2)}(\omega^{2}-k^{2})]\,.
\end{equation}
The physical interpretation of Eq.~(\ref{eq:pp_derivation5}) is straightforward~\cite{Watkins:1991zt, Falkowski:2012fb}: the classical background field configuration $\phi(z,t)$, describing the two approaching and colliding bubble walls, is decomposed via its Fourier transform into modes of definite four-momentum $\chi \equiv \omega^{2}-k^{2}$. Modes with $\chi>0$ behave as propagating off-shell quanta $\phi^{*}_{\chi}$ of the field $\phi$ with effective mass-squared $m^{2}=\chi$, and Eq.~(\ref{eq:pp_derivation5}) simply integrates, over all such modes contained in the collision, the number of  quanta of a given mass multiplied by their probability to decay into physical particles. From the optical theorem, the imaginary part of the 2-point 1PI Green function is given by the sum over the matrix element to all final states
\begin{equation}\label{eq:pp_derivation6}
\text{Im}[\tilde{\Gamma}^{(2)}(\chi)] = \frac{1}{2}\sum_{\alpha}\int d\Pi_{\alpha}|\mathcal{\overline{M}}(\phi^{*} \rightarrow \alpha)|^{2}\Theta(\chi - \chi_{\text{min},\alpha})\,,
\end{equation}
where $|\mathcal{\overline{M}}(\phi \rightarrow \alpha)|^{2}$ is the spin-averaged matrix element of the decay of $\phi$ to all possible final states $\alpha$, $\Theta$ is the Heaviside step function, and $\chi_{\text{min}} =  (\sum m_{\alpha})^{2}$ is the minimum energy required to produce the final state particles. Changing the variables to $\chi = \omega^{2}-k^{2}$ and $\xi =  \omega^{2}+k^{2}$ and integrating over $\xi$, we finally arrive at
\begin{equation}\label{eq:pp_rate}
\frac{N}{A} \approx\frac{1}{2\pi^{2}}\int_{\chi_{\text{min}}}^{\chi_\text{{max}}}d\chi f(\chi)\text{Im}\Big[\tilde{\Gamma}^{(2)}(\chi)\Big]\,,
\end{equation}
where
\begin{equation}\label{eq:eff_factor}
f(\chi) = \int_{\chi}^{2\chi_{\text{max}}-\chi}d\xi\frac{1}{\sqrt{\xi^{2}-\chi^{2}}}|\tilde{{\phi}}(\xi,\chi)|^{2}\,.
\end{equation}

Inspecting Eq.~(\ref{eq:pp_rate}), we can see that it expresses a sum over all particles that can be produced, multiplied by the probability of producing each. Specifically, $\text{Im}[\tilde{\Gamma}^{(2)}(\chi)]$ encodes the particle physics part, whereas $f(\chi)$ expresses the efficiency of particle production at scale $\chi$. $\chi_{\text{min}}$ is set either by the masses of the particles being produced, or by the Infrared (IR) cutoff set by the maximum size of the bubble at the collision $R_{*}$, i.e., $\chi_{\min}=\max\left[\left(\sum_i m_i\right)^2,R_*^{-2}\right]$, whereas $\chi_{\text{max}} = (\gamma_{w}/\ell_{w})^{2} \equiv \Lambda^{2}$,\footnote{In the numerical KG analysis, we do not impose $\chi_{\max}=\Lambda^{2}$ as a hard cutoff; instead, we evaluate the numerical spectrum over the finite $\chi$ range specified in Appendix~\ref{app:fitting}.} where $\Lambda$ is the inverse thickness of the boosted bubble wall. Thus, and as explained in Ref.~\cite{Mansour:2023fwj}, particle production from bubble collision involves two parts. The first part involves pure particle physics that is derived from the action, whereas the second involves the Fourier-transformed scalar field $\tilde{\phi}(k,\omega)$ that only involves the dynamics of spacetime during the phase transition that is independent of the decay itself. This split is a generic feature of the formalism, independent of the specific particle species being produced: as shown explicitly for scalars, fermions, and vector bosons coupled to the collision background in Ref.~\cite{Falkowski:2012fb}, changing the nature of the produced particle only ever modifies $\text{Im}[\tilde{\Gamma}^{(2)}(\chi)]$, typically altering its scaling with $\chi$ at high energies (roughly as $\chi^{0}$, $\chi^{1}$, and $\chi^{2}$ for scalars, fermions, and  vectors, respectively, in the case of direct tree-level couplings to $\phi$), while the efficiency factor $f(\chi)$ itself, which is the focus of this work, remains entirely determined by the collision dynamics of the background field. In~\cite{Mansour:2023fwj}, the efficiency factors $f(\chi)$ for both elastic and inelastic collisions were obtained in the limit $v_{w} \rightarrow 1$ by numerically solving the $SO(1,1)$ KG equation and the trapping equation, then fitting the results to obtain simple semi-analytic expressions. Our main task in this work is to obtain $f(\chi)$ over a broad range of $v_{w}$ by similar means.

\section{An attempted analytical solution for $v_{w} < 1$}
\label{sec:AttemptedAnalytical}
Before attempting to generalize the semi-analytical expressions to the $v_{w} < 1$ case numerically, it is worthwhile attempting to obtain a generalization of the original analytical expressions of the efficiency factors obtained in Ref.~\cite{Falkowski:2012fb}. There, the boosted bubble wall profiles $\phi(x,t)$ were modeled using boosted kink profiles with $\gamma_{w} \gg 1$ and with $v_{w}$ set to 1. The trapping equation~(\ref{eq:trapping}) was assumed to hold, and the KG equation was solved for the following generic potential~\cite{Mansour:2023fwj,Jinno:2019bxw}:
\begin{equation}\label{eq:Potential}
V(\phi) = av_{\phi}^{2}\phi^{2} - (2a+4)v_{\phi}\phi^{3} + (a+3)\phi^{4}\,.
\end{equation}
For $a>0$, this potential has a local minimum at $\phi=0$, corresponding to the false vacuum, and a global minimum at $\phi=v_\phi$, corresponding to the true vacuum. We define the degeneracy parameter $\epsilon$ as
\begin{equation}
\epsilon \equiv \frac{V_{\text{max}}-V(\phi=0)}{V_{\text{max}}-V(\phi=v_{\phi})} = \frac{a^{3}(a+4)}{a^{3}(a+4) + 16(a+3)^{3}}\,,
\end{equation}
where $V_{\text{max}}$ is the height of the potential barrier separating the true and false minima. It can be shown that $\epsilon$ increases the height of the potential barrier, which subsequently determines whether the collision is elastic or inelastic. More specifically, it was found in Ref.~\cite{Jinno:2019bxw} that $\epsilon > 0.22$ corresponds to elastic collisions, whereas $\epsilon < 0.22$ corresponds to inelastic collisions. The masses of $\phi$ in the false and true minima are given by
\begin{equation}\label{eq:masses}
m_{f}^{2} = \frac{d^{2}V(\phi)}{d\phi^{2}}\Big|_{\phi=0} = 2 a v_{\phi}^{2}\,, \hspace{5mm} m_{t}^{2} = \frac{d^{2}V(\phi)}{d\phi^{2}}\Big|_{\phi=v_{\phi}} = 2 (a+6)v_{\phi}^{2}\,.
\end{equation}

Before generalizing the efficiency factors to $v_w<1$, it is useful to recall the physical picture underlying this collision dynamics~\cite{Falkowski:2012fb}. Writing the equation of motion as the $(1+1)$-dimensional PDE $(\partial_{t}^{2}-\partial_{z}^{2})\phi = -dV/d\phi$, one notes that outside each bubble wall, where the field sits in either the true or false vacuum, the potential-gradient term vanishes identically, so the equation reduces there to the free wave equation. Since any function of $z\pm t$ solves the free equation, the incoming kink profiles continue to superpose essentially unperturbed as they approach one another, which is precisely the physical justification for the free-passage/trapping-equation ansatz of Eq.~(\ref{eq:trapping}). At the moment of collision itself, the field makes a brief excursion in field space to values beyond $v_\phi$ in the small region around the collision point, where $dV/d\phi\neq0$; the subsequent fate of this excursion, whether it is driven back over the barrier into the false vacuum (elastic) or remains trapped in the basin of the true vacuum (inelastic), is fixed entirely by the local shape of $V(\phi)$~\cite{Falkowski:2012fb}. This also makes clear why the free-passage approximation is expected to become progressively less reliable as $v_w$ decreases: away from $v_w\to1$, the potential-gradient term is no longer parametrically negligible relative to the kinetic term even {\it before} the walls reach the collision point, so the incoming profiles need not superpose as cleanly as assumed. We return to a quantitative assessment of this point later in the work (see Appendix~\ref{app:validity}).

To generalize the analytic factors in Ref.~\cite{Falkowski:2012fb} to $v_{w} <1$, we start with the same ansatz, namely, that of a static kink to model the bubble profiles
\begin{equation}\label{eq:static_kink}
\phi_{\text{kink}}(z) = \frac{v_{\phi}}{2}\Big[1+\tanh{\Big( \frac{z}{\ell_{w}}\Big)}\Big]\,,
\end{equation}
then we boost the left and right bubble walls with an arbitrary $v_{w}$
\begin{equation}
\phi_{L,R}(z,t) = \frac{v_{\phi}}{2}\Big[ 1 \pm \tanh{\Big( \Lambda [z \pm v_{w}t]\Big)}\Big]\,.
\end{equation}
Thus, under the free-passage superposition ansatz, the total pre-collision field profile is
\begin{equation}\label{eq:bubble_profile}
\phi(z,t<0) = v_{\phi} + \frac{v_{\phi}}{2}\Big[\tanh{\Big( \Lambda[z+v_{w}t]\Big)} - \tanh{\Big( \Lambda[z-v_{w}t]\Big)} \Big]\,.
\end{equation}
However, there is an important subtlety. The trapping equation~(\ref{eq:trapping}) assumed that the kinetic term of the KG equation dominates over the potential gradient, such that the profiles of colliding bubbles simply add up. This is a very good approximation for ultrarelativistic walls, however, it is less rigorous for $v_{w} < 1$. We will return to this point later on.

We begin by tackling the case of a Perfectly Elastic (PE) collision, but before we consider the bubble profiles with a finite thickness, let us consider bubbles with infinitely thin walls since this case will be useful later on. If the bubbles have infinitely thin walls, then their profile can be modeled by a step function
\begin{equation}\label{eq:step_func}
\phi_{\infty} = v_{\phi}\Theta(z^{2} - v_{w}^{2}t^{2})\,,
\end{equation}
which generalizes, to arbitrary $v_w$, the thin-wall profile $h_{\infty}$ used in Ref.~\cite{Falkowski:2012fb} strictly for $v_w\to1$ (their Eq.~(2.2)). Fourier-transforming Eq.~(\ref{eq:step_func}), changing the variables to $\xi$ and $\chi$, inserting in Eq.~(\ref{eq:eff_factor}), and performing the integral with $\xi_{\text{max}} = 2\Lambda^{2} - \chi$, we obtain
\begin{equation}\label{eq:PE_factor_step1}
f_{\text{PE}}(\chi) = \frac{16 v_{\phi}^{2}}{\chi^{2}}\Bigg[ \frac{1+v_{w}^{2}}{2v_{w}}\ln{\Bigg( \frac{1+v_{w}\sqrt{1-\chi/\Lambda^{2}}}{1-v_{w}\sqrt{1-\chi/\Lambda^{2}}}\Bigg)}  -\frac{(1-v_{w}^{2})\sqrt{1-\chi/\Lambda^{2}}}{1-v_{w}^{2}(1-\chi/\Lambda^{2})} \Bigg]\Theta(\Lambda^{2}- \chi)\,.
\end{equation}
Notice that in the limit $v_{w} \rightarrow 1$ we get
\begin{equation}\label{eq:PE_factor_step2}
f_{\text{PE}}(\chi) =\frac{16v_{\phi}^{2}}{\chi^{2}}\ln{\Bigg[ \frac{1+\sqrt{1-\chi/\Lambda^{2}}}{1-\sqrt{1-\chi/\Lambda^{2}}}\Bigg]}\,,
\end{equation}
which matches the expression in Ref.~\cite{Falkowski:2012fb}, derived there for the step-function profile $h_\infty$. To incorporate a finite wall thickness in the perfectly elastic case, we model the collision by reflecting the two finite-thickness walls at $t=0$. The incoming and outgoing configurations are therefore combined into the symmetric reflected-wall profile, i.e.,~\footnote{The direct generalization, keeping $v_w$ explicit rather than set to unity, of the finite-thickness tanh ansatz $h_{l_w}$ introduced in Ref.~\cite{Falkowski:2012fb} (their Eq.~(2.3))}
\begin{equation}\label{eq:bubble_profile_PE}
\phi(z,t) = \frac{v_{\phi}}{2}\Big[ \tanh{\Big(\Lambda[v_{w}t+|z|]\Big)} - \tanh{\Big(\Lambda[v_{w}t-|z|] \Big)}\Big]\,,
\end{equation}
which, following the same steps, yields
\begin{equation}\label{eq:PE_factor_finite}
f_{\text{PE}}(\chi) =\frac{8\pi^{2}v_{\phi}^{2}}{\Lambda^{2}}\int_{\chi}^{\infty} d\psi \frac{\psi + \chi}{\sqrt{\psi^{2}-\chi^{2}}}\frac{1}{[(1-v_{w}^{2})\psi +(1+v_{w}^{2})\chi]^{2}}\frac{1}{\sinh^{2}{\Big[ \frac{\pi}{2\sqrt{2}\Lambda v_{w}}\sqrt{\psi + \chi}\Big]}}\,,
\end{equation}
which when taking the limit $v_{w} \rightarrow 1$, matches Eq.~(2.24) in Ref.~\cite{Falkowski:2012fb}. The situation is a bit tricky for inelastic collisions. In this case, the bubble profile is given by Eq.~(\ref{eq:bubble_profile}) for $t < 0$, however, for $t > 0$ we need to evolve the bubble profile in the potential given in Eq.~(\ref{eq:Potential}) according to the KG equation. We will not repeat the calculation here and refer the interested reader to~\cite{Falkowski:2012fb} for details. It is sufficient to note that one only needs to change $\frac{\gamma_{w}}{\ell_{w}} \rightarrow \frac{v_{w} \gamma_{w}}{\ell_{w}} = v_{w} \Lambda$ in their bubble profile for the Totally Inelastic (TI) collision to obtain the generalized one
\begin{equation}\label{eq:TI_bubble_profile}
\phi_{\text{TI}}(z,t) =
\begin{cases}
v_{\phi} + \frac{v_{\phi}}{2}\Big[ \tanh{(\Lambda[z+v_{w}t])} - \tanh{(\Lambda[z-v_{w}t])}\Big], & t < 0,\\
v_{\phi}\Bigg[ 1+ \frac{v_{w}}{\Lambda} \int_{0}^{\infty}dp_{z} \frac{p_{z}\cos{(zp_{z})}}{\sqrt{p_{z}^{2}+m_{t}^{2}}\sinh{(\frac{\pi p_{z}}{2\Lambda})}} \sin{\Big(\sqrt{p_{z}^{2}+m_{t}^{2}}t)}   \Bigg], & t > 0\,.
\end{cases}
\end{equation}
Fourier transforming Eq.~(\ref{eq:TI_bubble_profile}) and following the same steps, we obtain the generalized efficiency factor for TI collisions
\begin{equation}\label{eq:TI_factor}
f_{\text{TI}}(\chi) = \frac{\pi^{2}v_{\phi}^{2}v_{w}^{2}}{2\Lambda^{2}} \frac{1}{(\chi-m_{t}^{2})^{2}}\int_{\chi}^{\infty} d\psi \frac{\psi-\chi}{\sqrt{\psi^{2}-\chi^{2}}} \frac{[(1-v_{w}^{2})(\psi-\chi)+2m_{t}^{2}]^{2}}{[(1-v_{w}^{2})\psi +(1+v_{w}^{2})\chi]^{2}}\frac{1}{\sinh^{2}{\Big[\frac{\pi}{2\sqrt{2}\Lambda}\sqrt{\psi - \chi}\Big]}}\,,
\end{equation}  
and it is an easy exercise to show that taking the limit $v_{w} \rightarrow 1$ reproduces the corresponding expression in Ref.~\cite{Falkowski:2012fb} (their Eq.~(2.25), up to the regulator prescription of their Eq.~(2.26), which we have not imposed here and discuss further in Sec.~\ref{sec:ShakyaFit}). As in Ref.~\cite{Falkowski:2012fb}, it is worth emphasizing that Eqs.~(\ref{eq:PE_factor_finite}) and~(\ref{eq:TI_factor}) encode purely the collision dynamics of the classical background field and make no reference to the species of particle eventually produced; the latter enters only through $\text{Im}[\tilde{\Gamma}^{(2)}(\chi)]$ in Eq.~(\ref{eq:pp_rate}), as discussed in Sec.~\ref{sec:Dynamics}.

\begin{figure}[!t] 
\centering
\includegraphics[width=\textwidth]{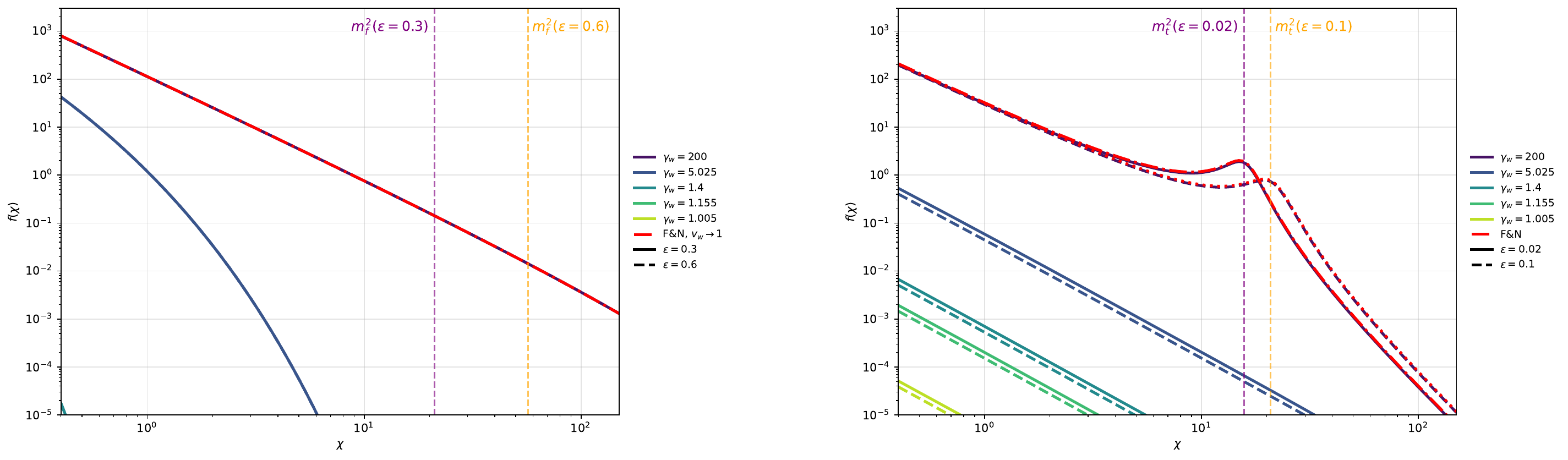}
\caption{The analytical efficiency factors for several benchmarks for $v_{w} <1$ compared to the runaway case $v_{w} \rightarrow 1$ (red) fixed at $\gamma_{w} = 200$ and taken from~\cite{Falkowski:2012fb} for both elastic (left) and inelastic collisions (right). For elastic collisions, the solid (dashed) line corresponds to $\epsilon =0.3\,(0.6)$, whereas for inelastic collisions the solid (dashed) line corresponds to $\epsilon =0.02\,(0.1)$. The vertical dashed lines in the elastic (inelastic) panel correspond to the squared mass of the  scalar field in the false (true) vacuum.}
\label{fig:AnalyticlComp}
\end{figure}

To show the impact of $v_{w} < 1$, we show in Figure~\ref{fig:AnalyticlComp} the efficiency factors in Eqs.~(\ref{eq:PE_factor_finite}) and~(\ref{eq:TI_factor}) for $\gamma_{w} = \{200, 5.025, 1.4, 1.155, 1.005\}$ corresponding to $v_{w} =\{ \rightarrow 1, 0.98, 0.7, 0.5, 0.1\}$, and we compare them to the expressions in Ref.~\cite{Falkowski:2012fb} which are only valid for $v_{w} \rightarrow 1$. For each case, we select two benchmark points corresponding to $\epsilon =0.3, \, 0.6$ for the elastic case, and $\epsilon =0.02, \, 0.1$ for the inelastic case. First, we can see from the plot that setting $v_{w} \rightarrow 1$ in the runaway regime is quite justified, as the efficiency factor is indistinguishable from that where the exact $v_{w}$ is used. Second and more importantly, the analytical expressions suggest that the efficiency of particle production drops drastically when we move away from the runaway regime. For instance, even when $v_{w} = 0.98$, the efficiency of particle production could drop by approximately 2-3 orders of magnitude. This picture, however, is misleading. As we shall see later on, these analytical expressions are not accurate and serve as poor generalizations to the $v_{w} < 1$ regime, precisely because the free-passage assumption underlying Eqs.~(\ref{eq:bubble_profile_PE}) and~(\ref{eq:TI_bubble_profile}) becomes increasingly unreliable away from $v_{w}\to1$, as anticipated above. We will address this issue in Subsection~\ref{subsec:AnalyticalAcc}.

\section{The numerical fit in the runaway regime}\label{sec:ShakyaFit}
\begin{figure}[!t] 
\centering
\includegraphics[width=\textwidth]{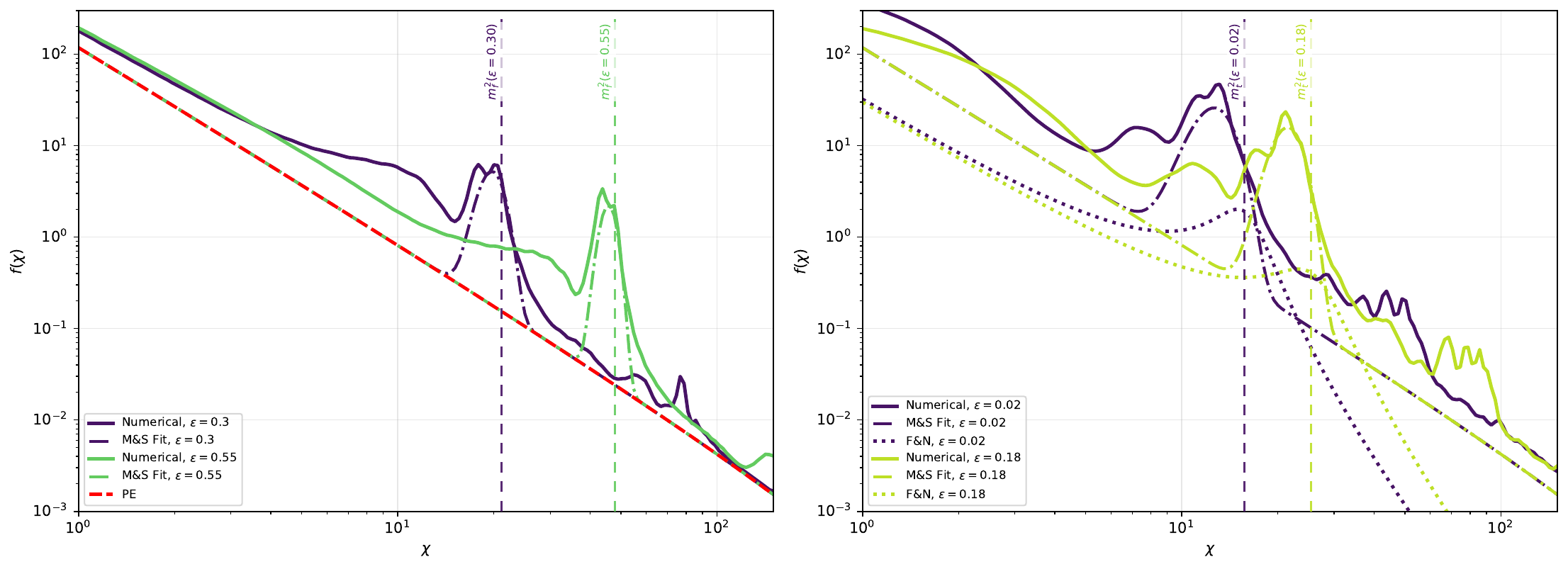}
\caption{A comparison between the numerical fit of the efficiency factors using Eqs.~(\ref{eq:MS_elastic}) and~(\ref{eq:MS_inelastic}) (dot-dashed) and the numerical solution of the $SO(1,1)$ KG equation (solid) for elastic (left) and inelastic (right) collisions. In addition, the PE solution from Eq.~(\ref{eq:PE_factor_step2}) (red dashed) and the analytical factors from Ref.~\cite{Falkowski:2012fb} (dotted) are shown on the elastic and inelastic cases respectively. We show two benchmarks corresponding to $\epsilon = \{0.3,0.55\}$ for the elastic case and $\epsilon = \{0.02,0.18\}$ for the inelastic one.}
\label{fig:MSComparisons}
\end{figure}
In this section, we will review the numerical fit of the efficiency factors in the runaway regime where $v_{w} \rightarrow 1$. These factors were obtained in Ref.~\cite{Mansour:2023fwj} to which we refer the interested reader for more details. The authors based their approach on the following four assumptions: 
\begin{enumerate}
\item \textbf{The trapping equation}: In the ultrarelativistic regime, the kinetic term of the KG equation is assumed to dominate over the potential gradient term and superposition applies. Thus, Eq.~(\ref{eq:trapping}) is assumed to hold,

\item \textbf{Light-front coordinates}: In the ultrarelativistic regime, the KG equation possesses an $SO(1,1)$ symmetry and one can use the light-front coordinate $s = \sqrt{t^{2} - x^{2}}$. Therefore, the KG equation simplifies to Eq.~(\ref{eq:KG_SO}),

\item \textbf{Bubble wall thickness}: In the ultrarelativistic regime, the bubble walls are assumed to be infinitely thin. Thus, the bubble walls are modeled by a step function as in Eq.~(\ref{eq:step_func}),

\item \textbf{Bubble interaction}: The two bubbles approach each other for $t< 0$ with $\phi = 0 $ between them and $\phi = v_{\phi}$ inside each bubble, then they collide at $(z,t) = (0,0)$. After the collision, the field configuration is evolved by solving the $SO(1,1)$ equation. In other words, the interaction is only turned on for $t\geq 0$.
\end{enumerate}

The authors perform their numerical analysis on an interval $[-L/2, L/2]$ in both $z$ and $t$. Discrete $N \times N$ points were generated in the $z$--$t$ space with separation $d = L/N$ to describe $\phi(z,t)$, then the discrete Fourier transform is obtained numerically and used to evaluate the efficiency factors using Eq.~(\ref{eq:eff_factor}). The default numerical choices are $L = 40 v_{\phi}^{-1}$ and $d = 0.01 v_{\phi}^{-1}$, with $d < 0.1 m_{f,t}^{-1}$. Finally, the authors used simple semi-analytical expressions to model the efficiency factors. 

What the authors of Ref.~\cite{Mansour:2023fwj} found is that the efficiency factors in both elastic and inelastic collisions can be approximated by a continuum given by the PE solution with a Gaussian bump on top of it. It is instructive to note that this bump is not an entirely novel feature: it is the numerical manifestation of a resonance that was already anticipated analytically in Ref.~\cite{Falkowski:2012fb}, where the totally-inelastic efficiency factor $f_{\rm TI}(\chi)$ was found to formally diverge as $\chi \to m_{h}^{2}$ (in our notation, $\chi\to m_t^2$), reflecting the fact that the idealized TI calculation, carried out over infinite spacetime and hence containing an infinite number of post-collision oscillation cycles, places unbounded spectral power exactly at the oscillation frequency. In Ref.~\cite{Falkowski:2012fb}, this divergence was regulated by hand through the substitution $(\chi-m_{h}^{2})^{2}\to (\chi-m_{h}^{2})^{2}+m_{h}^{6}\ell_{w}^{2}/\gamma_{w}^{2}$, motivated by the fact that the solution is not expected to remain valid over distances much larger than the bubble radius. The numerical fit of Ref.~\cite{Mansour:2023fwj} effectively supersedes this {\it ad hoc} regulator: by solving the field evolution over a finite, physically motivated spacetime volume rather than analytically over all of space and time, the numerical study automatically produces a finite, Gaussian-shaped peak in place of the formal divergence, and, crucially, reveals that the same oscillatory feature is also present, and non-negligible, for elastic collisions, a case for which the purely analytical treatment of~\cite{Falkowski:2012fb} provides no such resonance at all. Specifically, they found~\footnote{In Ref.~\cite{Mansour:2023fwj}, $m_{f}$ and $m_{t}$ are wrongly interchanged.}
\begin{align}
f_{\text{elastic}}(\chi) & = f_{\text{PE}}(\chi) + \frac{v_{\phi}^{2}L_{p}^{2}}{15m_{f}^{2}}\exp{\Bigg( \frac{-(\chi - m_{f}^{2} + 12m_{f}/L_{p})^{2}}{440m_{f}^{2}/L_{p}^{2}}\Bigg)}\,,\label{eq:MS_elastic}\\
f_{\text{inelastic}}(\chi) & = f_{\text{PE}}(\chi) + \frac{v_{\phi}^{2}L_{p}^{2}}{4m_{t}^{2}}\exp{\Bigg( \frac{-(\chi - m_{t}^{2} + 31m_{t}/L_{p})^{2}}{650m_{t}^{2}/L_{p}^{2}}\Bigg)}\,,\label{eq:MS_inelastic}
\end{align}
where $f_{\text{PE}}(\chi)$ is given by Eq.~(\ref{eq:PE_factor_step2}) and $L_{p} = \text{min}(R_{*},\Gamma^{-1})$, with $R_{*}$ being the bubble radius at the collision and $\Gamma$ being the decay width of $\phi$. The role played by $L_p$ here is directly analogous to the role of the bubble radius $R_B$ invoked in Ref.~\cite{Falkowski:2012fb} to justify their divergence-regulating prescription: both quantities set the maximal physical length (or, equivalently, time) scale over which the post-collision scalar-wave oscillations can coherently build up before the idealized single-collision treatment itself breaks down.

We plot Eqs.~(\ref{eq:MS_elastic}) and~(\ref{eq:MS_inelastic}) in Figure~\ref{fig:MSComparisons} together with the numerical solution of the $SO(1,1)$ equation. In addition, we show the PE solution for the elastic case and the analytical expressions obtained in Ref.~\cite{Falkowski:2012fb} for the inelastic case. Our results agree well with those in Ref.~\cite{Mansour:2023fwj}, where we see that the fit functions approximate the numerical solution to very good accuracy. We also find that the analytical expressions in Ref.~\cite{Falkowski:2012fb} tend to underestimate the efficiency factors, especially for $\chi \gg 1$.

As emphasized in Ref.~\cite{Falkowski:2012fb}, the efficiency factor $f(\chi)$ encodes exclusively the Fourier content of the classical collision background, independent of the species of particle ultimately produced; the fit functions of Eqs.~(\ref{eq:MS_elastic}) and~(\ref{eq:MS_inelastic}) can therefore be convolved with $\text{Im}[\tilde{\Gamma}^{(2)}(\chi)]$ for any particle species of interest, scalars, fermions, or vector bosons, via Eq.~(\ref{eq:pp_rate}), exactly as was done for the step-function/analytic case in Ref.~\cite{Falkowski:2012fb}. We caution, however, that whenever the resulting particle yield for a given channel approaches a sizable fraction of the energy stored in the colliding walls, $E_w/A$, backreaction effects neglected throughout this formalism can become important, and the results should be interpreted accordingly~\cite{Falkowski:2012fb}.

\section{Numerical solution of the KG equation for $v_{w} < 1$}
\label{sec:KGnumerical}
\begin{figure}[t!] 
\centering
\includegraphics[width=\textwidth]{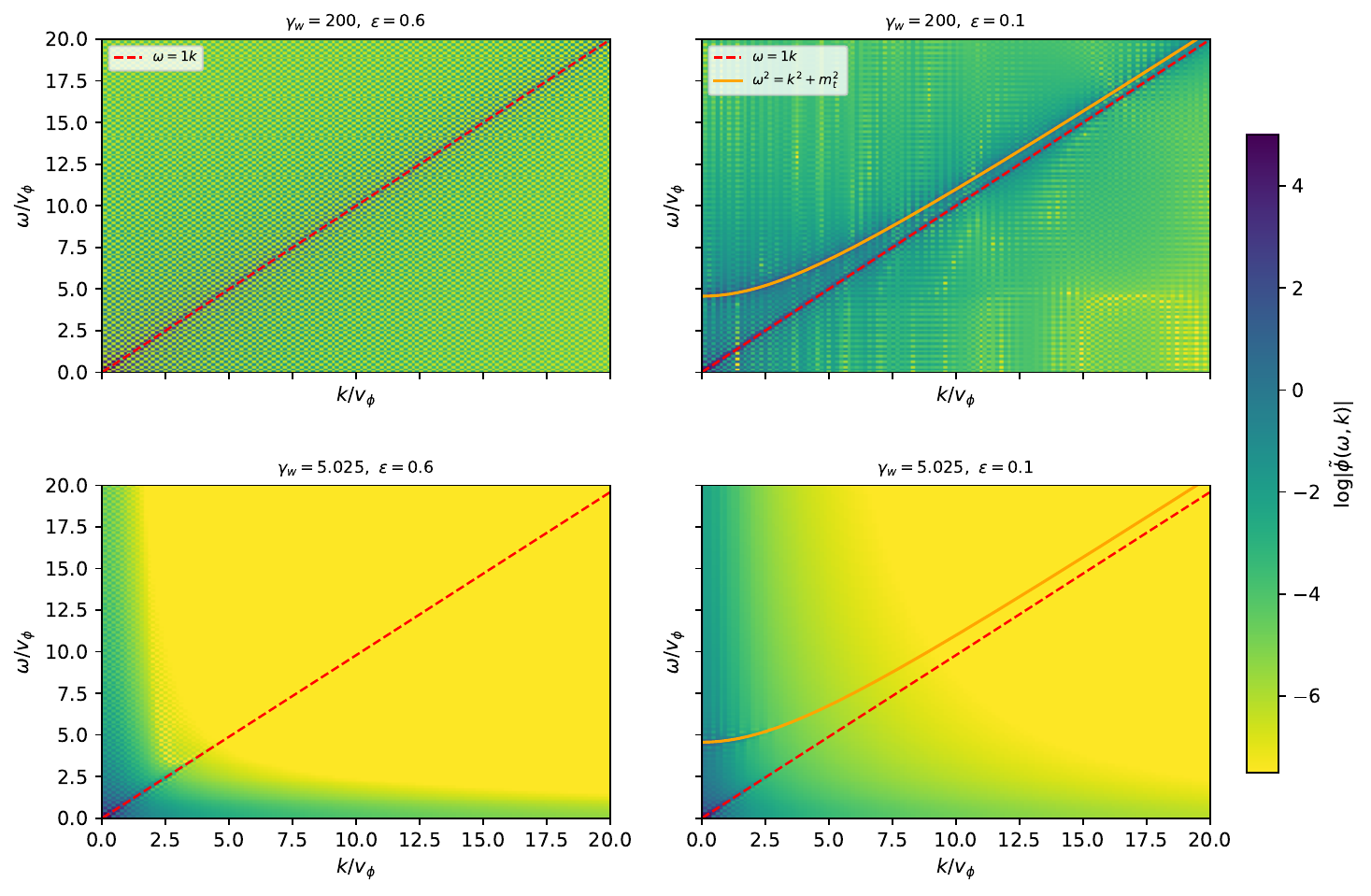}
\caption{A few examples of the numerical discrete Fourier transform. The top row is obtained by solving the $SO(1,1)$-symmetric KG equation for $\gamma_{w}=200$, assuming a step-function wall profile, whereas the bottom row is obtained by solving the full KG equation for $\gamma_{w}=5.025$, assuming walls with finite thickness. The left (right) plots are for $\epsilon = 0.6~(0.1)$ corresponding to elastic (inelastic) collisions.}
\label{fig:FourierProfilesSample}
\end{figure}

Both the analytical solution of Ref.~\cite{Falkowski:2012fb} and the semi-analytical one of Ref.~\cite{Mansour:2023fwj} are valid in the runaway regime where $v_{w} \rightarrow 1$. However, the results of the previous sections indicate that the runaway expressions cannot be reliably extrapolated to $v_{w}<1$. A dedicated numerical treatment is therefore required to determine the dependence of the particle-production efficiency factors on $v_{w}$. This is not merely a technical refinement: as discussed already in Ref.~\cite{Falkowski:2012fb}, whether a given bubble wall actually achieves the runaway, ultrarelativistic behavior assumed throughout Secs.~\ref{sec:Dynamics} and~\ref{sec:AttemptedAnalytical} is itself a dynamical question, set by the competition between the pressure driving the wall and the friction exerted on it by the surrounding plasma (or, in our more model-independent framework, by whatever mechanism eventually halts the wall's acceleration). Depending on the details of this friction, a bubble wall may either reach a stationary terminal velocity $v_w<1$ well before collision, or may continue accelerating with $\gamma_w$ growing throughout the expansion, reaching $v_w\to1$ only asymptotically; both outcomes are realized across the space of FOPT models, and the former is at least as generic as the latter~\cite{Falkowski:2012fb}. Therefore, generalizing the particle production efficiency factors requires careful treatment. In general, we will follow an identical route to Ref.~\cite{Mansour:2023fwj} in that we utilize a numerical solution of the KG equation and then use curve fitting to obtain generalized efficiency factors for any $v_{w}$. In addition, we will use an identical numerical setup as theirs to facilitate comparison. However, we update their main assumptions. Specifically,
\begin{itemize}
\item \textbf{The KG equation}: Away from the runaway regime, we cannot assume any $SO(1,1)$ symmetry. Thus, we cannot use any light-front coordinate and  we solve the KG equation as a partial differential equation in $t$ and $z$,

\item \textbf{Bubble wall thickness}: Away from the ultrarelativistic regime, the infinitely thin-wall approximation becomes unreliable. We therefore model the walls using the boosted kink profiles given by Eq.~(\ref{eq:bubble_profile}) before the collision and subsequently evolve $\phi(z,t)$ according to the full KG equation. This mirrors the distinction already noted in Ref.~\cite{Falkowski:2012fb} between the idealized step-function profile $h_\infty$ and the finite-thickness profile $h_{l_w}$: even in the ultrarelativistic limit, the latter was found to capture physical effects (the exponential damping of high-momentum Fourier modes) that the former misses entirely, and we expect this distinction to only grow more important as $v_w$ departs from unity and the wall can no longer be treated as parametrically thin on the timescale of the collision,

\item \textbf{The trapping equation}: As $v_{w}$ is lowered, the kinetic term in the KG equation becomes less dominant than the potential gradient term, which limits the validity of the trapping equation~(\ref{eq:trapping}). Nonetheless, we will continue to assume its validity for any $v_{w}$ to avoid significant complications in the numerical solution. We will discuss the accuracy of this assumption in Appendix~\ref{app:validity},

\item \textbf{Bubble interaction}: Accounting for the deformation of the walls caused by their interaction before the collision would substantially complicate the numerical treatment. We therefore prescribe the incoming walls as noninteracting boosted kink profiles for $t<0$, and evolve the field using the full KG equation from $t=0$ onward.
\end{itemize}

These four prescriptions together allow us to numerically track the full post-collision dynamics of the scalar field directly from the equation of motion $(\partial_t^2-\partial_z^2)\phi = -dV(\phi)/d\phi$, in the spirit of the direct numerical solutions of this same PDE performed in Ref.~\cite{Falkowski:2012fb} for representative choices of $\epsilon$ and $\gamma_w$, but here carried out systematically over the full two-dimensional grid of $(v_w,\epsilon)$ values needed to construct generalized efficiency factors, and without restricting to $v_w\to1$.

\begin{figure}[t!] 
\centering
\includegraphics[width=\textwidth]{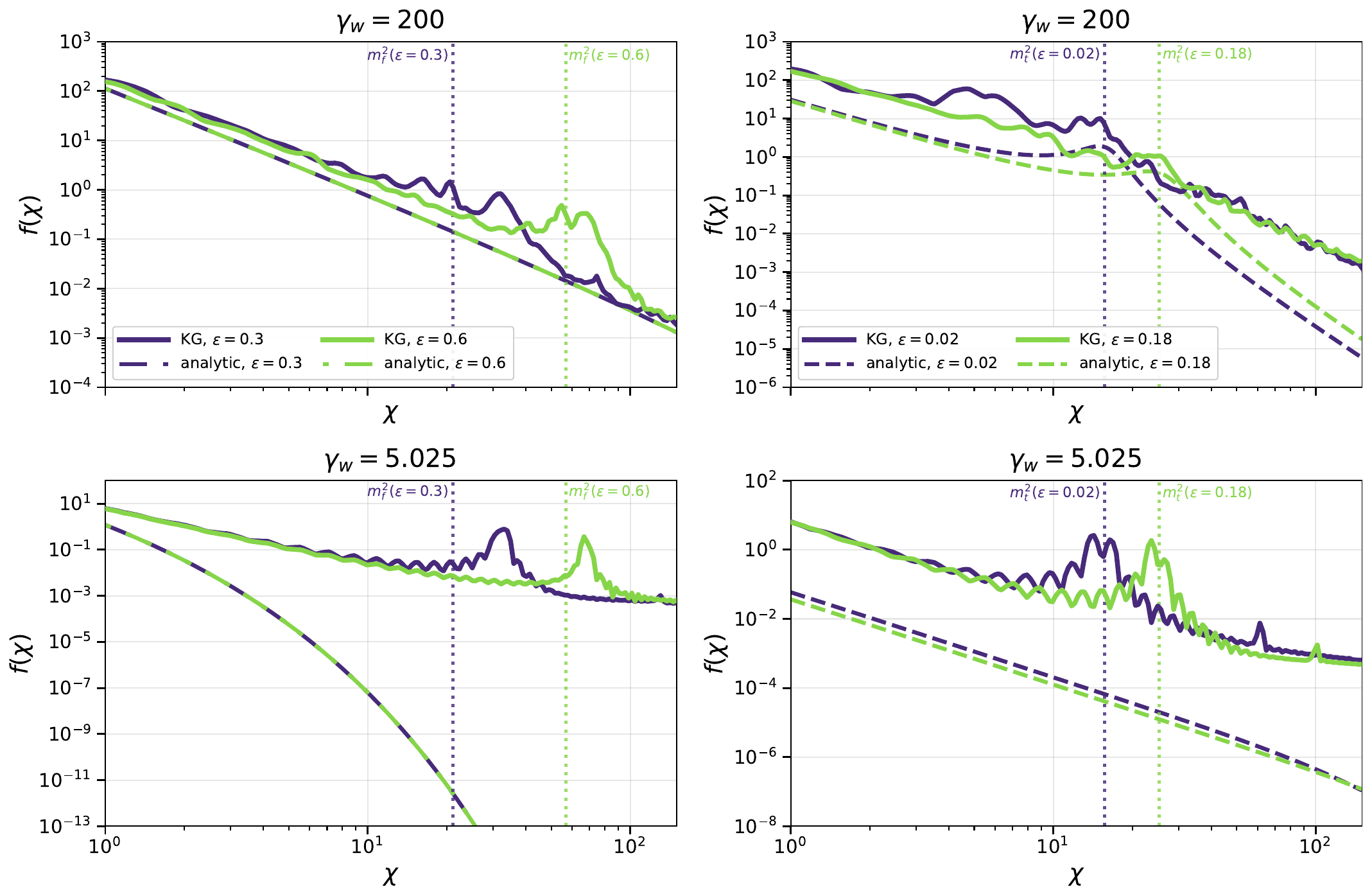}
\caption{Comparison of the analytical expressions of the efficiency factors obtained in Eqs.~(\ref{eq:PE_factor_finite}) and ~(\ref{eq:TI_factor}) (dashed) with the full numerical solution of the KG equation (solid) for $\gamma_{w} = 200$ (top row) and $5.025$ (bottom row), for elastic (left) and inelastic (right) collisions. For both elastic and inelastic collisions, we plot two benchmarks corresponding to $\epsilon = 0.3$ (purple) and $\epsilon = 0.6$ (green) for the elastic case, and $\epsilon = 0.02$ (purple) and $\epsilon = 0.18$ (green) for the inelastic case. The vertical dotted lines indicate $m^{2}_{f}$ and $m^{2}_{t}$.}
\label{fig:AnalyticVsKG}
\end{figure}

In order to find the numerical solution, we solve the full KG equation using the collision-centered initial conditions implied by the boosted-kink and trapping prescriptions, and we use the same numerical setup as the previous section. We then find the discrete numerical Fourier transform $\tilde{\phi}(k,\omega)$ (see Figure~\ref{fig:FourierProfilesSample} for a few examples) and solve the integral in Eq.~(\ref{eq:eff_factor}). After that, we perform curve-fitting on the numerical solution to find $f(\chi)$.

\subsection{Accuracy of the analytical generalization}
\label{subsec:AnalyticalAcc}
\begin{figure}[t!] 
\centering
\includegraphics[width=0.8\textwidth]{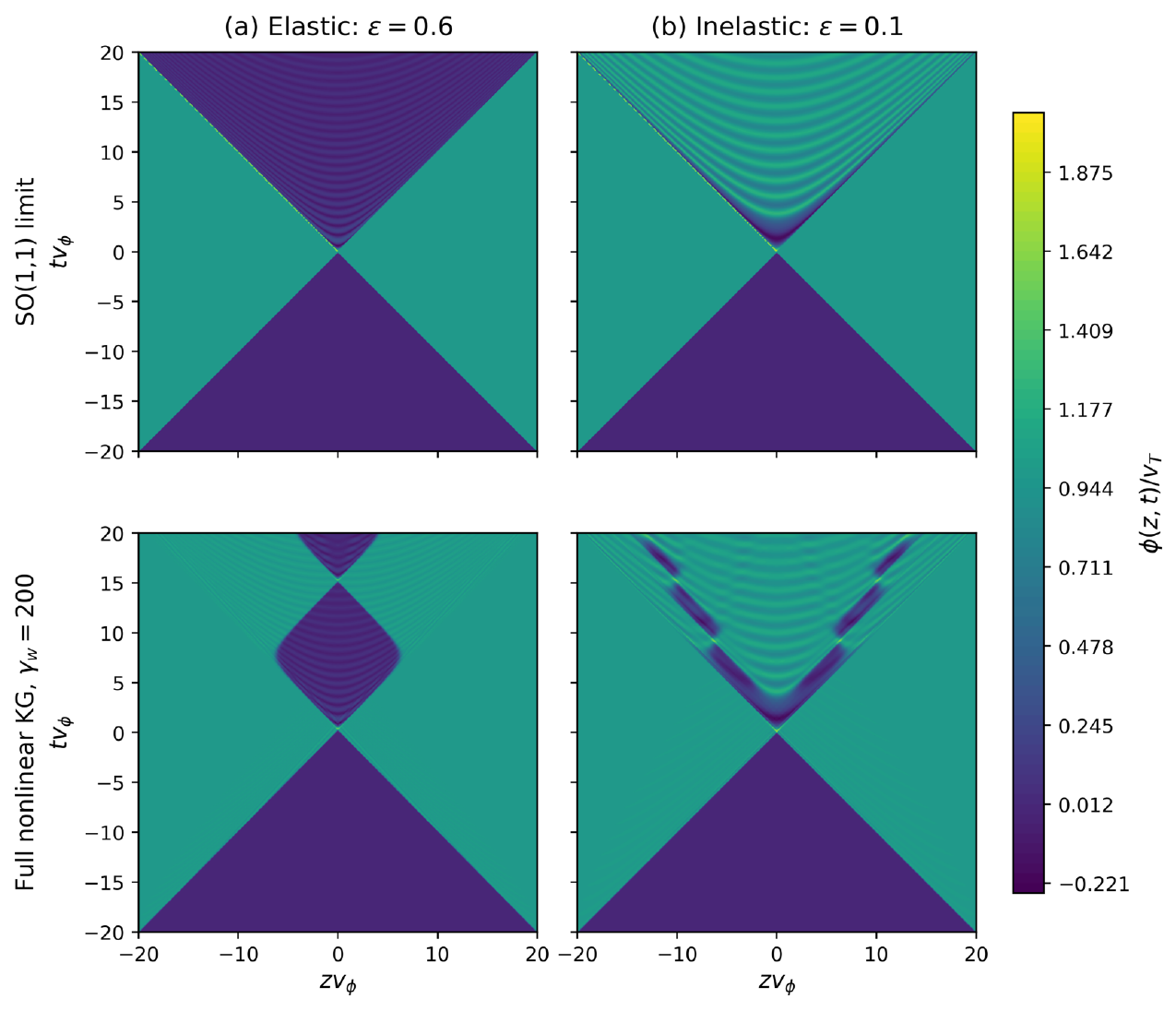}
\caption{Comparison of the scalar field spacetime configurations in elastic (left) and inelastic (right) collisions, when $\phi(z,t)$ is evolved according to the $SO(1,1)$ KG equation for $t>0$ (top) and when it is evolved according to the KG PDE starting from $t = -L/2$ (bottom). Here we set $\gamma_{w} = 200$. Notice that for our fit in Figure~\ref{fig:MSComparisons2}, we turn on the interaction for $t>0$.}
\label{fig:CollisionPlots}
\end{figure}
Before we describe the fitting, let us first investigate the accuracy of the analytical expressions of the efficiency factors we found for the non-runaway regime in Section~\ref{sec:AttemptedAnalytical} by comparing  Eqs.~(\ref{eq:PE_factor_finite}) and~(\ref{eq:TI_factor}) to the numerical solution of the full KG equation in Figure~\ref{fig:AnalyticVsKG}. As the plot shows, the analytical expressions have acceptable accuracy in the runaway regime; however, they generalize poorly to $v_{w} < 1$. For instance, for $\gamma_{w} = 5.025$ corresponding to $v_{w} = 0.98$, the analytical expression underestimates the actual numerical one by about 2 orders of magnitude in inelastic collisions, whereas for elastic collisions, it tends to underestimate it by about one order of magnitude for small $\chi$, with the accuracy deteriorating significantly for larger $\chi$. The accuracy becomes even worse as $v_{w}$ is lowered (see Appendix~\ref{app:results} for instance), which indicates that the analytical expressions in Eqs.~(\ref{eq:PE_factor_finite}) and ~(\ref{eq:TI_factor}) are only acceptable in the ultrarelativistic regime.

\subsection{Accuracy of the numerical fit in the runaway regime}
\label{subsec:NumericalAcc}
Here, we comment on the accuracy of the numerical fit in the runaway regime presented in Ref.~\cite{Mansour:2023fwj}. Specifically, we inspect the following aspects of the solution in Ref.~\cite{Mansour:2023fwj}: the assumption that the interaction is only turned on for $t>0$, the assumption of an $SO(1,1)$ symmetry which turns the KG equation from a PDE into the ODE in Eq.~(\ref{eq:KG_SO}), and the assumption that the bubbles can be modeled as  a step function instead of a boosted kink with a finite thickness.

To illustrate the impact of the first two assumptions, we show in Figure~\ref{fig:CollisionPlots} a comparison between the scalar field spacetime configurations for the solution of the $SO(1,1)$ equation with the potential turned on at $t>0$ and the solution of the full KG equation with the interaction starting at $-L/2$ for both elastic and inelastic collisions. As the plots show, turning on the interaction before the collision could have a significant impact on the spacetime configuration. We show a comparison for several $v_{w}$ benchmarks in Appendix~\ref{app:results}.

\begin{figure}[t!] 
\centering
\includegraphics[width=\textwidth]{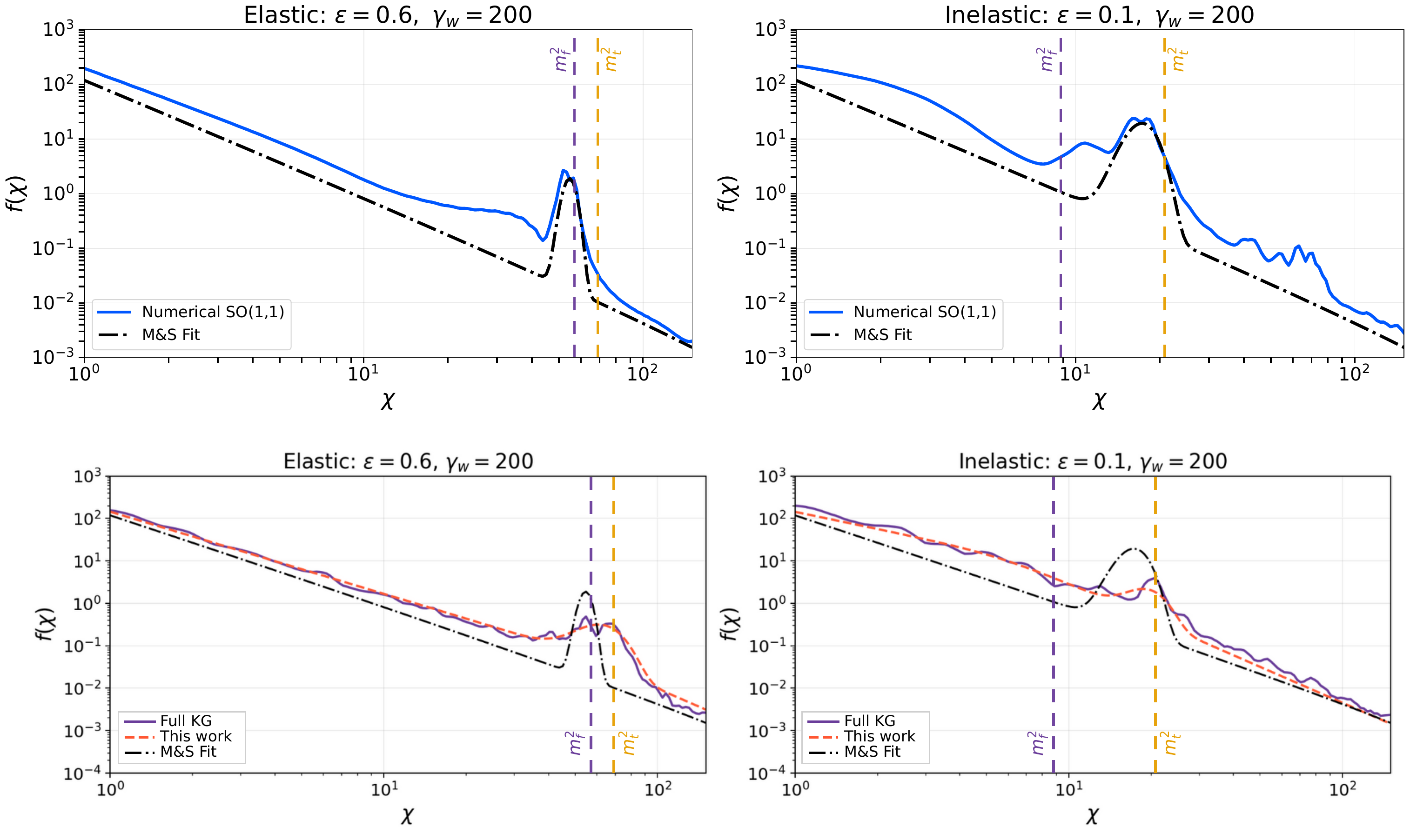}
\caption{(Top) the semi-analytical efficiency factors of Ref.~\cite{Mansour:2023fwj} given in Eqs.~(\ref{eq:MS_elastic}) and~(\ref{eq:MS_inelastic}) (black dot-dashed lines) superimposed on the numerical solution of the $SO(1,1)$ equation with the bubble profiles modeled by a step function, for both elastic collisions with $\epsilon = 0.6$ (left) and inelastic collisions with $\epsilon = 0.1$ (right). (Bottom) the same semi-analytical expressions compared to the numerical solution of the full KG equation with the bubble profile modeled by a boosted kink with a finite thickness (solid purple), together with the numerical fit obtained herein (red dashed). Here we fix $\gamma_{w} = 200$ as in Ref.~\cite{Mansour:2023fwj} and we show the locations of $m_{f}^{2}$ and $m_{t}^{2}$.}
\label{fig:MSComparisons2}
\end{figure}

Next, we investigate the impact of the bubble profile on the solution. To illustrate this, we compare the numerical fits of Ref.~\cite{Mansour:2023fwj} given in Eqs.~(\ref{eq:MS_elastic}) and~(\ref{eq:MS_inelastic}) with the numerical fit we obtain herein (see below) in Figure~\ref{fig:MSComparisons2}. Specifically, in the top row of the figure, we show the numerical solution of the $SO(1,1)$ equation with the bubble profile modeled as a step function, together with the numerical fit of Ref.~\cite{Mansour:2023fwj}, where we see that the fits reproduce the continuum and the bump fairly accurately within $\mathcal{O}(1)$ in both the elastic and the inelastic cases, although they miss some details such as a smaller bump in the inelastic case. Things are different, however, when we compare these fits to the full solution of the KG equation, with the bubble profile modeled as a boosted kink, as shown in the bottom row of the Figure. There, we see that although the efficiency factors of Ref.~\cite{Mansour:2023fwj} still reproduce the continuum part fairly accurately, it nonetheless does not reproduce the bump with good accuracy. Specifically, it tends to misplace the location of the bump and it tends to exaggerate its height. 

\begin{figure}[t!] 
\centering
\includegraphics[width=0.85\textwidth]{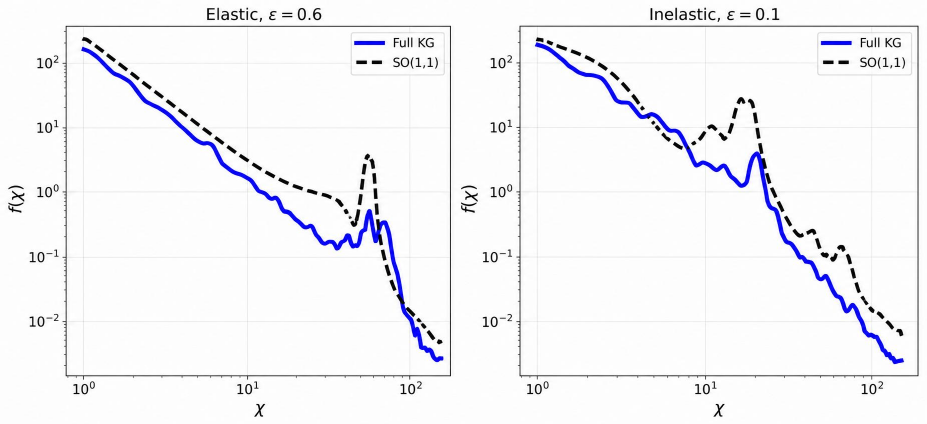}
\caption{A comparison between the numerical solution of the $SO(1,1)$ equation with a step function ansatz (dashed black), and the numerical solution of the full KG equation with a finite-thickness kink ansatz (solid blue) for $\epsilon = 0.6$ corresponding to elastic collisions (left) and $0.1$ corresponding to inelastic collisions (right).}
\label{fig:SOvsKG}
\end{figure}

For comparison, we show on the same plot the numerical fit we obtained in this work by solving the full KG equation and assuming a boosted kink with a finite wall thickness. There we see that it fits both the continuum and the bump better. We illustrate how we obtain this solution below. But before we move on to the numerical fit, let us clarify the shortcomings of the numerical fit of Ref.~\cite{Mansour:2023fwj} more concretely by comparing the solution of the $SO(1,1)$ equation with a step function ansatz with the solution of the full KG equation with a kink of finite thickness ansatz. We show that in Figure~\ref{fig:SOvsKG}, fixing $\gamma_{w} = 200$ and $\epsilon = 0.6$ and $0.1$ corresponding to elastic and inelastic collisions. 

There, we clearly see that the former tends to both misplace and exaggerate the height of the bump. This indicates that even in the ultrarelativistic regime, assuming an $SO(1,1)$ symmetry and modeling the bubble walls as infinitely thin step functions might not be too accurate, echoing, in the efficiency-factor language used throughout this work, the same lesson drawn in Ref.~\cite{Falkowski:2012fb} when comparing their thin-wall result $f_\infty(\chi)$ (Eq.~(\ref{eq:PE_factor_step2})) against the finite-thickness result $f_{l_w}(\chi)$ (Eq.~(\ref{eq:PE_factor_finite})): the finite thickness of the wall is not merely a formal refinement but can materially alter the shape of the efficiency factor, independent of how close $v_w$ is to unity.

\subsection{The numerical fit}
Inspecting the full KG solutions shown for example in Figures~\ref{fig:AnalyticVsKG} and~\ref{fig:MSComparisons2}, we find that similar to what was pointed out in Ref.~\cite{Mansour:2023fwj}, the efficiency factors obtained from solving the full KG equation can be modeled as a continuum with a Gaussian bump superimposed on it. Although the actual numerical solution is more complicated than that, exhibiting several other smaller bumps and oscillatory behavior, the major characteristics of these solutions can be modeled as such, i.e., they can be thought of as a continuum part and a Gaussian bump on top of it. The input physical parameters are
\begin{equation}
\{ v_{w}(\gamma_{w}),v_{\phi}, m_{f}, m_{t}, \ell_{w}, R_{*}, \Gamma \}\,,
\end{equation}
and the full function is given by
\begin{equation}\label{eq:FitFunc}
f_{X}(\chi) = \frac{v_{\phi}^{2}}{m_{X}^{4}}A_{X}\Big( \frac{\chi}{m_{X}^{2}}\Big)^{-p_{X}}\Bigg[ 1+\Big( \frac{\chi}{m_{X}^{2}x_{c,X}}\Big)^{4} \Bigg]^{-\Delta p_{X}/4} + \frac{v_{\phi}^{2}L_{p}^{2}}{m_{X}^{2}}B_{X}\exp{\Bigg[ -\frac{(\chi-m_{X}^{2}+c_{X}m_{X}/L_{p})^{2}}{2(d_{X}m_{X}/L_{p})^{2}}\Bigg]}\,,
\end{equation}
where $X \in \{E,I\}$ refers to elastic and inelastic collisions respectively, $m_{X} = m_{f} (m_{t})$ for elastic (inelastic) collision, and the fit constants $\{A_{X}, B_{X}, c_{X}, d_{X}, p_{X},\Delta p_{X},x_{c,X}\} $ are functions of $v_{w}$ and $\epsilon$ for fixed $v_{\phi}\ell_{w}$. We describe the fitting procedure in detail in Appendix~\ref{app:fitting} and provide numerical values of these fit constants in Appendix~\ref{app:constants}. The center and width of the Gaussian bump are determined by
\begin{align}
\chi_{\text{peak},X} & = m_{X}^{2}- \frac{c_{X}m_{X}}{L_{P}}\,\label{eq:chi_pleak},\\
\sigma_{\chi,X} & = \frac{d_{X}m_{X}}{L_{P}}\,.\label{eq:chi_sigma}
\end{align}

For the numerical catalogue used to determine the fit parameters in this work,
we fix $L_p = 40\,v_\phi^{-1}$, corresponding to the temporal extent of the numerical collision region.
Accordingly, the tabulated fit parameters in Appendix~\ref{app:constants} are calibrated for this choice. The explicit $L_p$ dependence in Eqs.~(\ref{eq:FitFunc})--(\ref{eq:chi_sigma}) retains the physical scaling motivated by Ref.~\cite{Mansour:2023fwj}, where $L_p=\min(R_*,\Gamma^{-1})$.

As in the original PE/TI decomposition of~\cite{Falkowski:2012fb}, the power-law piece of Eq.~(\ref{eq:FitFunc}) is tied to $m_f$ for elastic and $m_t$ for inelastic collisions, reflecting the mass scale of the oscillations that develop once the field settles (or is trapped) in the corresponding vacuum; our fit, however, allows the exponents $p_X$ and $\Delta p_X$, together with the Gaussian bump parameters $A_X$, $B_X$, $c_X$, $d_X$, and $x_{c,X}$, to vary with $v_w$, thereby capturing the smooth interpolation between the ultrarelativistic limit discussed in Sec.~\ref{sec:AttemptedAnalytical} and the increasingly suppressed, non-trivial behavior found at lower wall velocities.

\begin{figure}[t!] 
\centering
\includegraphics[width=\textwidth]{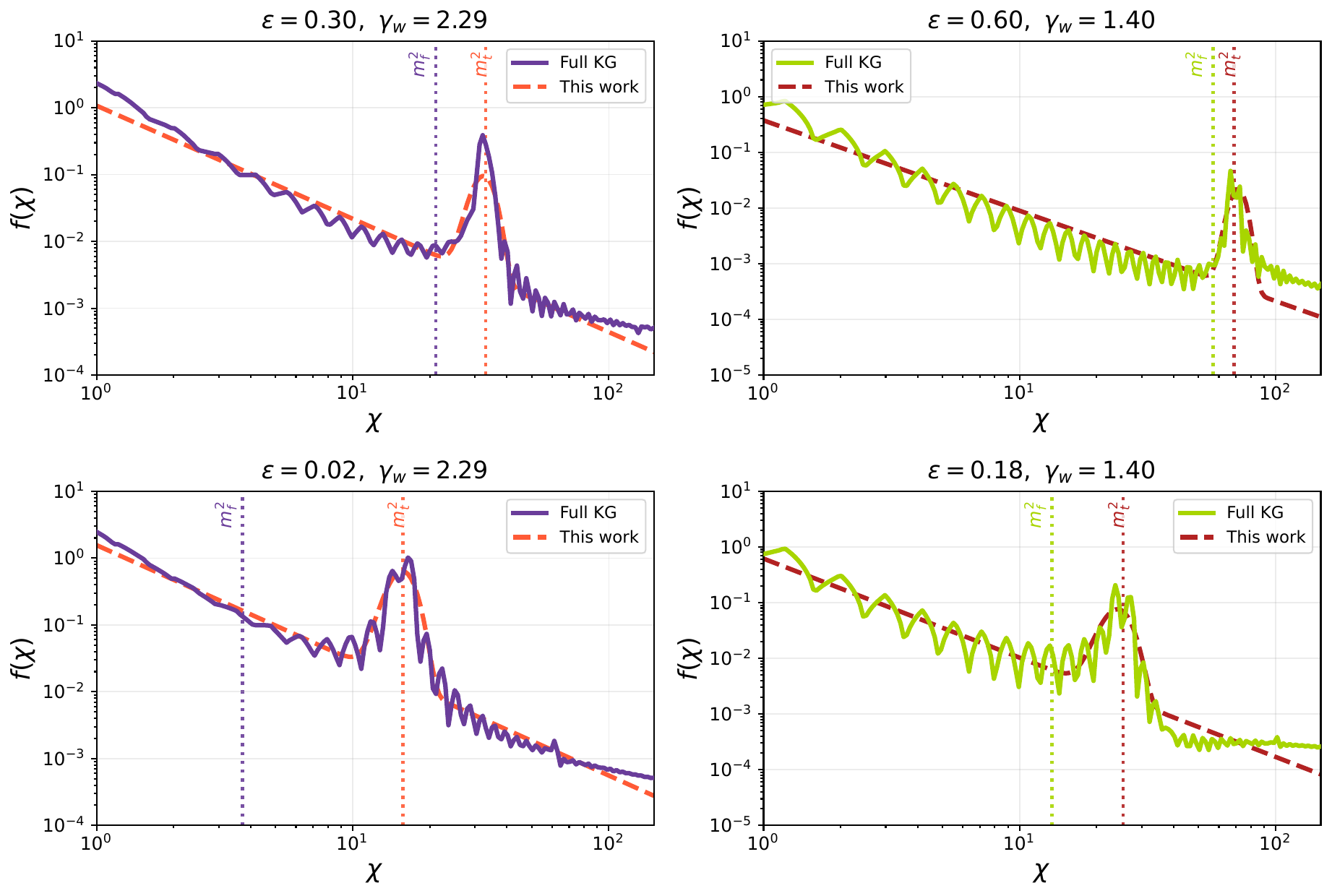}
\caption{Comparison of the numerical efficiency factors obtained from the full KG equation (solid) with the fit in Eq.~(\ref{eq:FitFunc}) (dashed) for several benchmark points. The top and bottom rows correspond to elastic and inelastic collisions, respectively. The vertical lines indicate the locations of $m_f^2$ and $m_t^2$.}
\label{fig:FitFunc}
\end{figure}

In Figure~\ref{fig:FitFunc}, we show the fit function in Eq.~(\ref{eq:FitFunc}) for several benchmark points of $\gamma_{w}$ and $\epsilon$ corresponding to both elastic and inelastic collision. Inspecting the plots, we find that the semi-analytical expression herein fits both the continuum part and the bump reasonably well, however, we note that when solving the full KG equation, the oscillatory nature of the KG equation becomes manifest, as evidenced by the small oscillations on both the continuum and the bump. Nonetheless, these oscillations have reasonably small amplitudes. In addition, we find that the accuracy of the fit becomes lower for larger $\chi$. Despite these residual oscillatory features, Eq.~(\ref{eq:FitFunc}) provides a useful parametrization of the numerical spectra throughout the benchmark grid. For the local fits entering the catalogue, the multiplicative RMS error ranges from approximately $1.14$ to $2.71$, with a median value of about $2.03$. The accuracy generally deteriorates toward the edges of the fitted spectral range, and the fit should therefore be interpreted as capturing the dominant continuum and resonance structures rather than as a percent-level representation of every oscillatory feature.

\section{Impact on particle production}
\label{sec:pp}
Having analyzed the impact of $v_w$ on the efficiency factors in the non-runaway regime, we next illustrate their impact on particle production within the conventional quadratic, two-point off-shell prescription reviewed in Sec.~\ref{sec:Dynamics}. The absolute particle yields obtained in this section should therefore be interpreted within the limitations of this formalism, which we
discuss in detail in Sec.~\ref{sec:Issues}.  To this end, we consider the same toy model as in Ref.~\cite{Mansour:2023fwj}, introducing a real scalar singlet $\psi$ coupled to $\phi$ through
\begin{equation}\label{eq:lagrangian}
\mathcal{L}_{\psi} = \frac{1}{2}(\partial \psi)^{2} - \frac{1}{2}m_{0}^{2}\psi^{2} - \frac{1}{2}\lambda_{\psi}v_{\phi} \phi\psi^{2}\,,
\end{equation}
with $\lambda_{\psi} = 1$, and $m_{0} = m_{\psi,f}$ being the bare mass in the false vacuum. Consequently, the mass in the true vacuum is $m_{\psi,t}^{2} = m_{0}^{2} + \lambda_{\psi} v_{\phi}^{2}$. The one-loop 2-point function is given by
\begin{equation}\label{eq:2pt}
\text{Im}\Gamma^{(2)} (\chi) = \frac{\lambda_{\psi}^{2}v_{\phi}^{2}}{32\pi}\sqrt{1-\frac{4m_{\psi}^{2}}{\chi}}\Theta(\chi - 4m_{\psi}^{2})\,,
\end{equation}
where here $m_{\psi} =m_{\psi,t}$. Strictly speaking, for $\lambda_\psi=1$ and a non-tachyonic scalar in the false vacuum, $m_0^2\geq0$ implies $m_\psi/v_\phi\geq1$. Following the conventional treatment of Ref.~\cite{Mansour:2023fwj}, we nevertheless show the two-point production kernel over a broader range of $m_\psi/v_\phi$ in order to illustrate its dependence on the production threshold. The region $m_\psi/v_\phi<1$ should therefore be regarded as a formal continuation of the kernel rather than as a consistent realization of the minimal Lagrangian in Eq.~(\ref{eq:lagrangian}). This restriction does not affect
the efficiency factors derived in the previous sections.

We emphasize that the efficiency factor $f(\chi)$ characterizes the collision background, whereas the absolute particle yield obtained after convolution with $\operatorname{Im}\widetilde{\Gamma}^{(2)}(\chi)$ is specific to the quadratic off-shell prescription adopted here.

\begin{figure}[t!]
\centering
\includegraphics[width=0.98\textwidth]{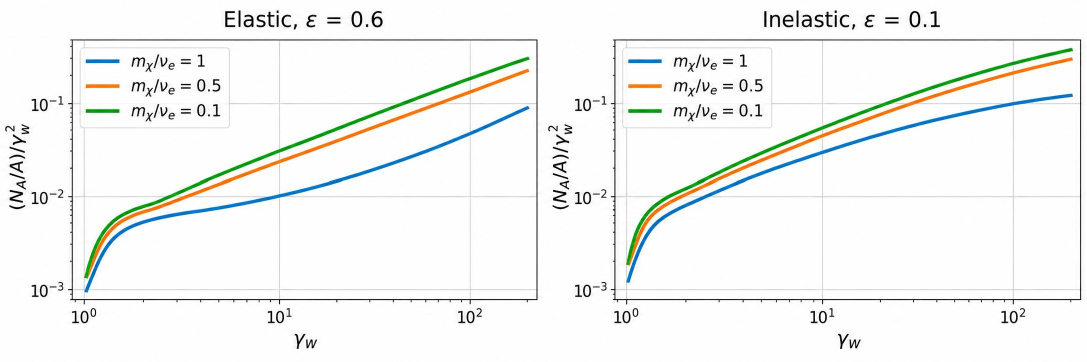}
\vspace{3mm}
\includegraphics[width=0.98\textwidth]{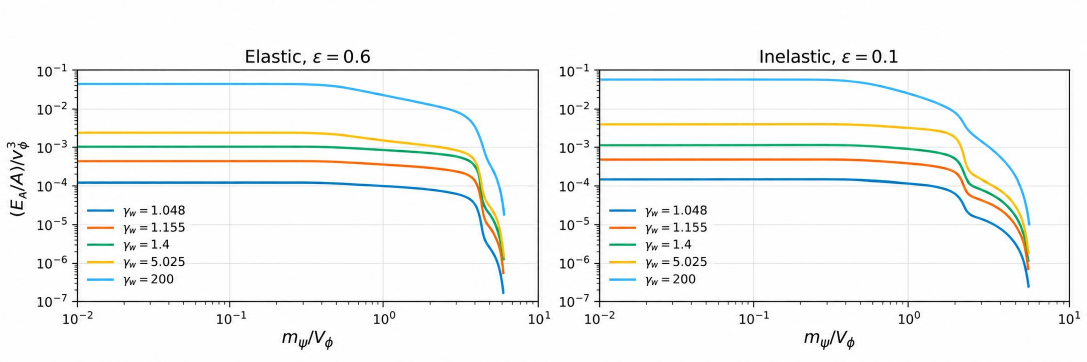}
\caption{(Top) impact of $v_{w}$ on the rate of particle production from bubble collisions. The plots show $(N_{\psi}/A)/v_{\phi}^{2}$ vs. $\gamma_{w}$ for several benchmark values of $m_{\psi}/v_{\phi}$. (Bottom) the impact of the mass of the produced particle on the production rate for several benchmark values of $\gamma_{w}$. On the left, $\epsilon = 0.6$ corresponding to elastic collisions, whereas on the right,  $\epsilon = 0.1$ corresponding to inelastic collisions.}
\label{fig:rateVgamma}
\end{figure}

We note that this choice of a scalar singlet directly coupled to $\phi$ is the simplest possible case, for which $\text{Im}\Gamma^{(2)}(\chi)$ scales as $\chi^{0}$ at large $\chi$. As shown in Ref.~\cite{Falkowski:2012fb} for the runaway regime, a direct Yukawa coupling to a fermion instead leads to $\text{Im}\Gamma^{(2)}(\chi)\propto \chi$, while a direct coupling to a vector boson leads to $\text{Im}\Gamma^{(2)}(\chi)\propto \chi^{2}$; since the efficiency factor $f(\chi)$ itself is universal and depends only on the collision dynamics (see Sec.~\ref{sec:Dynamics}), the qualitative impact of $v_w$ on the production rate that we discuss below for scalars is expected to translate directly to fermions and vector bosons, with the overall rate further enhanced for the latter two species due to the more favorable high-$\chi$ scaling of $\text{Im}\Gamma^{(2)}$. This can be convolved with the relevant efficiency factor in Eq.~(\ref{eq:FitFunc}) to find the number of particles produced per unit area by numerically solving the integral in Eq. (\ref{eq:pp_rate}). We show the impact of $v_{w}$ on the rate of particle production by plotting $(N_{\psi}/A)/v_{\phi}^{2}$ against $\gamma_{w}$ for several values of $m_{\psi}/v_{\phi}$ for both elastic and inelastic collisions in the top panel of Figure~\ref{fig:rateVgamma}. As one would expect, increasing the wall velocity significantly enhances the rate of particles produced whether the collision is elastic or inelastic. Within the numerical prescription adopted here, this dependence arises through the change of the collision spectrum $f(\chi)$ itself as $v_w$ and $\gamma_w$ are varied. In particular, larger boosts produce a harder and more efficient collision spectrum. Although the inverse boosted wall thickness $\Lambda=\gamma_w/\ell_w$ defines the characteristic physical ultraviolet scale associated with the wall profile, we do not impose $\chi_{\max}=\Lambda^2$ as a hard cutoff in the numerical KG calculation; instead, the numerical spectrum is evaluated over the finite $\chi$ range specified in Appendix~\ref{app:fitting}.
 
In particular, we find that when $\gamma_{w}$ is not much larger than 1, the suppression in the particle production rate becomes severe. Another observation one can glean from the plots is that the rate is enhanced when the particles produced are lighter. This is intuitive as more particles can be produced for a given energy budget if the species is lighter. More specifically, the phase space factor $\sqrt{1-4m_{\psi}^{2}/\chi}$ becomes larger when $m_{\psi}$ decreases, which makes the integration region wider. The impact of the mass is shown in the bottom panel of Figure~\ref{fig:rateVgamma} where the same behavior is observed. We can observe from the bottom plots that the rate becomes strongly suppressed as
the pair-production threshold $4m_\psi^2$ moves toward the upper end of the numerically resolved spectrum. Once the threshold lies above the numerical $\chi$ range used in the calculation, the corresponding integral vanishes. This should not be interpreted as a fundamental kinematic cutoff of the collision, but rather as the limit of the spectral range probed in the present numerical analysis.

We can observe from the bottom plots that the rate becomes suppressed when the energy available from the collision is not sufficient to produce particles that heavy.

\begin{figure}[t!]
\centering
\includegraphics[width=\textwidth]{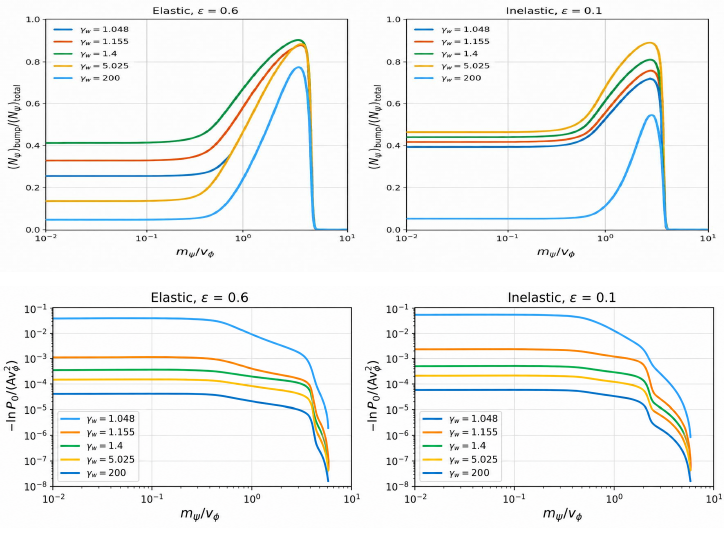}
\caption{(Top) the fraction of particles produced by the bump relative to the total number produced by the collision for several benchmark points in $\gamma_{w}$. (Bottom) the normalized vacuum-decay exponent $-\ln{P_{0}/(A v_{\phi}^{2}})$ for several benchmark points of $\gamma_{w}$. On the left, $\epsilon = 0.6$ corresponding to elastic collisions, whereas on the right,  $\epsilon = 0.1$ corresponding to inelastic collisions.}
\label{fig:BumpFraction}
\end{figure}

An important characteristic of the efficiency factors is the existence of a bump. This observation was one of the main results of Ref.~\cite{Mansour:2023fwj}. Therefore, this observation deserves more consideration. In the top row of Figure~\ref{fig:BumpFraction}, we show the fraction of the particles produced from the bump relative to the total number produced as a function of the particle mass. We show this for multiple benchmarks of $\gamma_{w}$. We observe from the plot that a significant fraction of the particles are produced from the bump, with the fraction being larger for smaller $\gamma_{w}$ (although this is not a one-to-one correspondence). Another important feature is that as $m_\psi$ increases such that the pair-production threshold $\chi_{\rm th}=4m_\psi^2$ approaches the position of the oscillation-induced bump, $\chi\simeq m_X^2$, where $m_X=m_f$ ($m_t$) for elastic (inelastic) collisions, the fraction produced by the bump rises significantly and can become dominant. Equivalently, this occurs when $2m_\psi\simeq m_X$. Near this threshold the available phase space is strongly suppressed, so the spectral weight concentrated in the bump becomes increasingly important.~\footnote{The fraction does not reach unity because when the pair-production threshold $\chi_{\rm th}=4m_\psi^2$ coincides with the center of the bump, $\chi\simeq m_X^2$, the phase-space factor vanishes,
\[
\operatorname{Im}\widetilde{\Gamma}^{(2)}
\propto
\sqrt{1-\frac{4m_\psi^2}{\chi}}
\;\longrightarrow\;0.
\]
Equivalently, the coincidence occurs for $2m_\psi\simeq m_X$.}

Next, we analyze the probability of particle production. To this end, we show the normalized vacuum-decay exponent $-\ln{P_{0}/(A v_{\phi}^{2}})$, which is related to the vacuum persistence, in the bottom row of Figure~\ref{fig:BumpFraction}. Vacuum persistence essentially expresses the probability of NOT producing particles. Specifically, the probability of vacuum persistence can be expressed as
\begin{equation}\label{eq:vacPersistence}
P_{0} = |\langle 0_{\text{in}}  |0_{\text{out}} \rangle|^{2} = e^{-2\text{Im}\Gamma}\,.
\end{equation}
Thus when $P_{0} \simeq 1$ there are essentially no particles produced, whereas $P_{0} \ll 1$ indicates strong particle production. In Figure~\ref{fig:BumpFraction}, we show $-\ln{P_{0}/(A v_{\phi}^{2}})$, which is the vacuum-decay exponent per unit collision area normalized by $v_{\phi}^{2}$, against $m_{\psi}/v_{\phi}$. We observe that increasing $\gamma_{w}$ lowers vacuum persistence, i.e., increases the probability of particle production. In addition, vacuum persistence is enhanced as the particle's mass increases. This is consistent with our findings above.

\begin{figure}[t!]
\centering
\includegraphics[width=\textwidth]{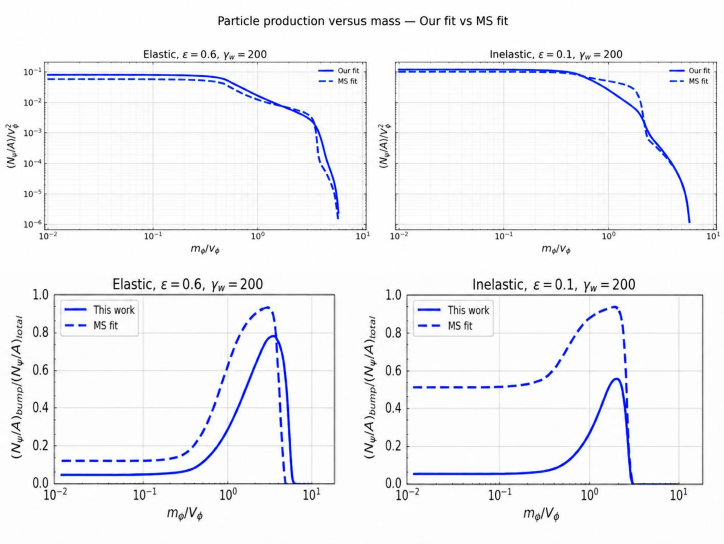}
\caption{A comparison of the results of this work (solid) and those of Ref.~\cite{Mansour:2023fwj} (dashed) in the runaway regime with $\gamma_{w} = 200$. The top plot shows the normalized vacuum-decay exponent $-\ln{P_{0}/(A v_{\phi}^{2}})$ and the bottom shows the fraction of the particles produced by the bump.}
\label{fig:MSComparison3}
\end{figure}

At this point, it is worthwhile comparing particle production obtained from the efficiency factor in Eq.~(\ref{eq:FitFunc}) with that in Ref.~\cite{Mansour:2023fwj} in the runaway regime. In the top row of Figure~\ref{fig:MSComparison3}, we show the vacuum-decay exponent, where we see that the results we obtain herein are very close to those in Ref.~\cite{Mansour:2023fwj}. Things change, however, when we compare the fraction of the particles produced by the bump, which is shown in the bottom row. As the plots show, the results in Ref.~\cite{Mansour:2023fwj} tend to overestimate the fraction of the particles produced by the bump, especially in inelastic collisions. This is consistent with the finding in Figures~\ref{fig:MSComparisons2} and ~\ref{fig:SOvsKG} above. 

Throughout this section we have neglected the backreaction of particle production on the dynamics of the background field $\phi(z,t)$ itself, treating $\psi$ as a spectator field in the presence of the classical collision source, as is standard in this formalism~\cite{Watkins:1991zt,Falkowski:2012fb,Mansour:2023fwj}. This treatment is self-consistent provided the energy radiated into $\psi$ quanta remains a small fraction of the kinetic energy stored in the colliding walls, $E_w/A \sim v_\phi^2\gamma_w/\ell_w$ (see Sec.~\ref{sec:Dynamics}); as shown explicitly in Ref.~\cite{Falkowski:2012fb} for direct couplings to vector bosons and, to a lesser extent, fermions, this condition can be violated for sufficiently strong portal couplings, in which case backreaction becomes important and the actual particle yield should be understood as effectively capped by $E_w/A$ rather than by the naive integral of Eq.~(\ref{eq:pp_rate}). For the modest scalar coupling $\lambda_\psi = 1$ considered here, we treat $\psi$ as a spectator field and neglect its backreaction on the collision background. The resulting particle-production rates should therefore be interpreted within the perturbative regime in which the energy transferred to the produced particles remains small compared with the energy stored in the colliding walls. This caveat becomes particularly important when extrapolating our results, or the underlying efficiency factors of Eq.~(\ref{eq:FitFunc}), to more strongly coupled scenarios or to species such as vector bosons whose production is parametrically more efficient.

\section{Limitations of the off-shell formalism}
\label{sec:Issues}
According to Ref.~\cite{Ghoshal:2026pew}, the treatment described in Section~\ref{sec:Dynamics} is problematic for two reasons. The first concerns the approximate collision profile underlying $f_{\text{PE}}$ in Eq.~(\ref{eq:PE_factor_step2}) which is related to off-shell efficiency factors, where one assumes that the two colliding walls reflect off one another. This assumption is problematic because for $\gamma_{w} \gg 1$, the two colliding walls overlap for a very short period of time $\sim \ell_{0}/ \gamma_{w}$ that is much shorter than the time scale over which the potential can impact them. Thus, it was argued in Ref.~\cite{Ghoshal:2026pew} that the two walls initially pass through each other with little interaction, similar to spectator partons in high-energy hadronic scattering.

The second and more serious issue raised in Ref.~\cite{Ghoshal:2026pew} concerns the approximation in Eq.~(\ref{eq:pp_derivation3}). Specifically, while Eq.~(\ref{eq:pp_derivation1}) is correct, approximating the full action with the 2-point function is problematic since the latter is an off-shell quantity, which introduces spurious gauge-dependence. More specifically, the off-shell two-point function is not invariant under field redefinitions. In gauge theories this also appears as gauge dependence, but the field-reparametrization problem is more general and can affect scalar and fermion production as well. 

As a remedy to these two issues, Ref.~\cite{Ghoshal:2026pew} introduced a different approach that they dubbed the ``on-shell approximation", which is analogous to the Weizs\"{a}cker-Williams equivalent-photon approximation in QED and to the parton model for high-energy hadron collisions. The on-shell approximation is intended for the hard regime, with invariant energies well above the wall-scalar mass, whereas coherent KG evolution remains relevant for soft modes.

Here we briefly comment on the impact of these two issues on our results. We delay a more rigorous treatment to future work. With regard to the first issue, we point out that as we are solving the full KG equation and extracting the efficiency factors by direct numerical fitting, the first issue is partially alleviated in our treatment. The second issue, however, is more problematic, as approximating the full action with the truncated 2-point function as in Eq.~(\ref{eq:pp_derivation3}) indeed introduces field-reparametrization dependence and gauge dependence. Although this truncation does not affect the derived efficiency factor per se, which is the main result in this work, it could nonetheless impact the rate of particle production. A formally invariant treatment would require the complete effective action evaluated consistently on a solution of the corresponding quantum equations of motion. Merely adding a finite number of higher-point terms to the two-point truncation is not sufficient.

This, however, runs into difficulty for two reasons. First of all, calculating the full action is difficult in general. Second, and perhaps more important, even with the full action, one cannot use the efficiency factor derived herein as is, since the production rate derived in Eq.~(\ref{eq:pp_rate}), which led to the efficiency factor in Eq.~(\ref{eq:eff_factor}), was derived by factorizing the 2-point function. More specifically, the full action contains higher-point terms
\begin{equation}\label{eq:FullAction}
\Gamma[\phi] = \Gamma^{(2)} +  \Gamma^{(3)} +  \Gamma^{(4)} + \cdots\,,
\end{equation}
and a decomposition similar to Eq.~(\ref{eq:pp_rate}) for the full action would require convolutions with several background momenta and their interference. This, in general, cannot be expressed as a single factor $f(\chi)$ and could be intractable. Thus, investigating the impact of approximating the full effective action by its two-point contribution deserves more scrutiny; we treat this issue in future work.

It is worth noting that the field-reparametrization issue discussed above is not the only limitation inherent to this formalism. As emphasized already in the context of the runaway regime in Ref.~\cite{Falkowski:2012fb}, the entire calculation treats the produced quanta as a perturbative source on top of a fixed classical background $\phi(z,t)$, neglecting the backreaction of particle production on the evolution of the background field itself; as we discussed in Sec.~\ref{sec:pp}, this approximation is only self-consistent when the energy radiated into particles remains small compared to the energy stored in the colliding walls, $E_w/A$, and can break down for sufficiently large couplings or for particle species, such as vector bosons, whose production is parametrically more efficient~\cite{Falkowski:2012fb}. Unlike the field-reparametrization issue, which is a conceptual shortcoming of the off-shell two-point truncation itself, the backreaction problem is in principle a solvable numerical one, requiring a self-consistent, coupled evolution of the background field and the produced particle degrees of freedom; we regard this, together with the field-reparametrization issue discussed above, as among the most pressing open directions for placing calculations of particle production from bubble collisions on a fully rigorous footing.

\section{Summary and Discussion}
\label{sec:Discussion}
In this work, we extended the study of particle production from FOPT bubble collisions beyond the runaway limit. By solving the full Klein-Gordon equation in the planar approximation with finite-thickness wall profiles, we found that straightforward analytical extensions of the ultrarelativistic results become unreliable as the wall velocity decreases. In particular, they can underestimate the numerical efficiency factors by several orders of magnitude. More concretely, the generalized analytical expressions of Eqs.~(\ref{eq:PE_factor_finite}) and~(\ref{eq:TI_factor}) are acceptable for $\gamma_{w}=200$, but already at $\gamma_{w}=5.025$ ($v_{w}=0.98$) they underestimate the numerical efficiency factor by about two orders of magnitude for inelastic collisions, and by about one order of magnitude at small $\chi$ for elastic collisions, with the discrepancy growing at larger $\chi$ and for smaller $v_{w}$. The apparent drastic drop of the efficiency away from $v_{w}\to 1$ suggested by the analytical expressions is therefore largely an artifact of the free-passage assumption, and the true suppression is considerably milder. Nevertheless, the principal spectral features identified in earlier ultrarelativistic studies remain present: a smooth continuum associated with the collision and a prominent bump generated by post-collision field oscillations, centered near $m_{f}^{2}$ ($m_{t}^{2}$) for elastic (inelastic) collisions. We provided a semi-analytical parametrization, Eq.~(\ref{eq:FitFunc}), consisting of a broken power-law continuum plus a Gaussian resonance peak, which describes both components over the range of wall velocities ($0.1\leq v_{w}<1$, i.e. $1.005\lesssim\gamma_{w}\leq 200$) and potential parameters ($0.02\leq\epsilon\leq 0.98$) studied. The fit constants are tabulated in Appendix~\ref{app:constants}, and for the benchmarks considered, the multiplicative RMS error ranges from approximately $1.14$ to $2.71$, with a median value of about $2.03$.

Revisiting the runaway regime, we also found that the simplifying assumptions of Ref.~\cite{Mansour:2023fwj}, namely the $SO(1,1)$ symmetry, the step-function wall profile and the late switching-on of the interaction, are not innocuous even at $\gamma_{w}=200$. While their fit reproduces the continuum well, it tends to misplace the location of the bump and to overestimate its height when compared with the full KG solution with finite-thickness walls.

As an illustrative application within the conventional two-point off-shell formalism, we applied these efficiency factors to a scalar-production toy model and found that particle production is strongly enhanced as $v_{w}$ increases and is suppressed for heavier particles because of the increasing kinematic threshold and phase-space suppression. The enhancement reflects the increasing efficiency and hardness of the collision spectrum as the wall boost is increased. The inverse boosted wall thickness $\Lambda=\gamma_w/\ell_w$ sets a characteristic physical ultraviolet scale of the wall profile, but it is not imposed as a hard upper cutoff on $\chi$ in our numerical calculation. For light particles the rate is essentially independent of the mass, while it drops sharply as the pair-production threshold moves into the high-$\chi$ tail of the collision spectrum and eventually outside the numerical
spectral range considered here. The vacuum-persistence analysis confirms this picture: the vacuum-decay exponent $-\ln P_{0}/(Av_{\phi}^{2})$ increases with $\gamma_{w}$ and decreases with the particle mass. The oscillation-induced bump can make a significant contribution to the total rate, particularly when the production threshold approaches the characteristic scalar mass governing the post-collision oscillations. In the ultrarelativistic benchmark ($\gamma_{w} =200$), our total production rates are close to those obtained in Ref.~\cite{Mansour:2023fwj}, although their treatment generally predicts a larger bump contribution, especially for inelastic collisions. Since the efficiency factor is universal, these qualitative conclusions are expected to carry over to fermions and vector bosons, for which the more favorable high-$\chi$ scaling of $\text{Im}\,\Gamma^{(2)}$ further enhances the rate.

Several limitations should be emphasized. Our numerical treatment remains planar and employs the trapping prescription, whose validity becomes increasingly uncertain away from the free-passage regime. Indeed, the criterion $\mathcal{C}_{\rm FP}\ll1$ derived in Appendix~\ref{app:validity} is not satisfied by any of the benchmarks considered, including $\gamma_{w}=200$ (for which $\mathcal{C}_{\rm FP}\simeq 0.13$--$0.49$), and would require $\gamma_{w}\gtrsim 236$--$463$ to be met; this should be kept in mind when interpreting the results, even though the numerical efficiency factors are still well described by our fit. A fully self-consistent calculation should evolve initially separated walls through the collision without imposing the trapping condition and should eventually incorporate bubble curvature, multiple successive collisions in the elastic case, and plasma effects. In addition, the backreaction of particle production on the background field is neglected, which is justified for the modest coupling $\lambda_{\psi}=1$ considered here but may fail for strongly coupled scenarios or for vector bosons. More fundamentally, the efficiency factor derived here is a robust characterization of the Fourier content of the classical collision background, but its interpretation as a universal particle-production kernel relies on truncating the effective action at quadratic order. As emphasized in Ref.~\cite{Ghoshal:2026pew}, the resulting off-shell prescription is field-reparametrization dependent and may parametrically overestimate hard production. The impact of the off-shell treatment will be examined in a future work.

\section*{Acknowledgements}
F. A. would like to thank SOKN Engineering for their support. A.A. is funded by the University of Sharjah via the HEP RG operational grant. 
S. M. gratefully acknowledges financial support from the Department of Science and Technology (DST), Government of India, under the INSPIRE Fellowship (No. DST/INSPIRE Fellowship/2020/IF200540), which supported the development of this work. The work of N.O. is supported in part by the United States Department of Energy Grant, No.~DE-SC0012447.  {\it The use of AI disclosure}: AI assistance (ChatGPT and Gemini) was used to generate code and plots, perform cross-checks and validation, and polish the language. However, all results were verified by the authors, who take full responsibility for the content.

\appendix

\section{The validity of the trapping equation}\label{app:validity}
According to~\cite{Giblin:2010bd}, if $\phi_{\text{FP}}$ is the solution corresponding to the free-pass regime where the trapping equation applies, then we can write the exact solution as $\phi = \phi_{\text{FP}} + \sigma$, where Ref.~\cite{Giblin:2010bd} estimates
\begin{equation}\label{eq:sigma}
\sigma \sim \frac{\ell_{w}^{2}|V'(2v_{\phi})|}{\gamma_{w}^{2}-1}\,.
\end{equation}
Thus, one can define the criterion for the validity of the FP regime by demanding that the exact solution be close to the FP solution, i.e,
\begin{equation}
\mathcal{C}_{\text{FP}} \equiv \frac{|\sigma|}{|\Delta\phi_{\text{FP}}|} \ll 1\,,
\end{equation}
where $ \Delta\phi_{\text{FP}} = 2v_{\phi}$. For the potential we consider in Eq.~(\ref{eq:Potential}), this translates to the following condition
\begin{equation}\label{eq:criterion}
\mathcal{C}_{\text{FP}} = \frac{2m_{t}^{2} + m_{f}^{2}}{v_{w}^{2}\Lambda^{2}} \ll 1 \,.
\end{equation}
Another way to look at this is to consider the Lorentz-contracted overlap time of the bubbles, which is given by
\begin{equation}
	t_{\text{overlap}} \sim \frac{\ell_{w}}{\gamma_{w}v_{w}}\,.
\end{equation}
Noting that the potential changes the field by $2v_{\phi}$ on a timescale of 
\begin{equation}
t_{V} \sim \sqrt{\frac{2v_{\phi}}{{V'(2v_{\phi})}}} =\frac{1}{\sqrt{2m_{t}^{2} + m_{f}^{2}}}\,,
\end{equation}
and the criterion for the validity of the FP regime can be expressed as the requirement that the overlap time be much faster than the time it takes the potential to change, i.e.,
\begin{equation}
\mathcal{C}_{\text{FP}} \simeq \Big( \frac{t_{\text{overlap}}}{t_{V}}\Big)^{2} \ll 1\,
\end{equation}
which leads to Eq.~(\ref{eq:criterion}). We calculate $\mathcal{C}_{\text{FP}}$ for a few benchmark points in Table~\ref{tab:FP_validity}. As the table shows, none of these benchmarks points (including $\gamma_{w} = 200$) is truly under control, which weakens our assumption of using the trapping equation. However, the numerical efficiency factors still fit the numerical solution reasonably well. A proper description of the non-runaway regime, however, would require more accurate treatment of the trapping equation. Specifically, the assumption that the potential gradient term in the KG equation is subleading compared to the kinetic term needs to be reevaluated. This is beyond the scope of this work.

\begin{table}[t]
\centering
\begin{tabular}{c|cccc}
\hline\hline
Case
& $C_{\rm FP}(\gamma_w=1.4)$
& $C_{\rm FP}(\gamma_w=5.025)$
& $C_{\rm FP}(\gamma_w=200)$
& Calibrated $\gamma_{\rm crit}$ \\
\hline
$\epsilon=0.60$
& $2.03\times 10^{4}$
& $8.02\times 10^{2}$
& $0.486$
& $463$ \\
$\epsilon=0.10$
& $5.26\times 10^{3}$
& $2.08\times 10^{2}$
& $0.126$
& $236$ \\
\hline\hline
\end{tabular}
\caption{The calculated $\mathcal{C}_{\text{FP}}$ for a few benchmarks. For each case, we also calculate the calibrated $\gamma_{\text{crit}}$, which represents the minimum value where the criterion in Eq.~(\ref{eq:criterion}) is met}
\label{tab:FP_validity}
\end{table}

\section{The fitting procedure}\label{app:fitting}
Here we describe the fitting procedure in more detail. We used the numerical setup described in Section~\ref{sec:ShakyaFit} and we set $v_{\phi} = 1$ and $\ell_{w} = 10 v_{\phi}^{-1}$ for the sake of numerical evaluation. With this choice, the scaling of the various physical quantities is shown in Table~\ref{tab:vphi_units}, whereas the fit constants are dimensionless. Values that lie on a fitting boundary indicate weak determination of the corresponding individual parameter, rather than a failure of the reconstructed efficiency factor.

\begin{table}[htbp]
    \centering
    \begin{tabular}{ccc}
        \toprule
        Quantity & variable & Scaling \\
        \midrule
        Scalar field
        & $\phi$
        & $v_{\phi}$ \\

        Mass
        & $m$
        & $v_{\phi}$ \\

        Length or time
        & $L$
        & $v_{\phi}^{-1}$ \\

        Momentum or frequency
        & $k,\,\omega$
        & $v_{\phi}$ \\

        Invariant mass squared
        & $\chi$
        & $v_{\phi}^{2}$ \\

        Potential or energy density
        & $V$
        & $v_{\phi}^{4}$ \\

        Efficiency factor
        & $f$
        & $v_{\phi}^{-2}$ \\
        \bottomrule
    \end{tabular}
     \caption {Scaling of the various numerical variables and physical quantities in units of $v_{\phi}$.}
      \label{tab:vphi_units}
\end{table}

The fitting function for both the elastic and inelastic cases is given in Eq.~(\ref{eq:FitFunc}), and the (transformed) fit constants can be cast into the vector 
\begin{equation}
\theta_{X}(\epsilon, \gamma_{w}) = (\ln{A},p,\ln{\Delta p}, \ln{x_{c}},\ln{B},c, \ln{d})_{X}\,.
\end{equation}
Inside a certain (elastic or inelastic) grid cell defined by $\epsilon_{i} \leq \epsilon \leq  \epsilon_{i+1}$ and $\gamma_{j} \leq  \gamma_{w} \leq \gamma_{j+1}$, we define
\begin{equation}
\alpha = \frac{\epsilon - \epsilon_{i}}{\epsilon_{i+1}-\epsilon_{i}}\,, \hspace{5mm} \beta = \frac{\ln{\gamma_{w}} - \ln{\gamma_{j}}}{\ln{\gamma_{j+1}}-\ln{\gamma_{j}}}\,,
\end{equation}
and the mathematical parameter fit is the piecewise bilinear surface
\begin{equation}
\widehat{\theta}_{r,X}(\epsilon,\gamma_{w}) = (1-\alpha)(1-\beta)\theta^{X}_{r,i,j} + \alpha(1-\beta)\theta^{X}_{r,i+1,j} + (1-\alpha)\beta\theta^{X}_{r,i,j+1} + \alpha\beta\theta^{X}_{r,i+1,j+1}\,.
\end{equation}

The interpolation is performed separately within the elastic and inelastic branches and never across the boundary at $\epsilon=0.22$. Thus, for $\epsilon<0.22$ we interpolate only among inelastic benchmark points, whereas for $\epsilon>0.22$ we interpolate only among elastic benchmark points. At the
borderline value $\epsilon=0.22$, the branch must be specified explicitly. To avoid confusion, we neglect this borderline value in the tabulated constants in Appendix~\ref{app:constants}. No extrapolation outside the corresponding tabulated branch domain is performed.

Exponentiation produces $A, \Delta p, x_{c}, B$, and $d$, whereas $p$ and $c$ are obtained directly. Thus, $p$ and $c$ are interpolated bilinearly, whereas the remaining parameters are interpolated geometrically. Before the collision, the two bubbles are modeled with boosted kinks with a finite thickness
\begin{equation}
\phi (z,t) = v_{\phi} + \frac{v_{\phi}}{2}\Big[ \tanh{\Big(\Lambda(z + v_{w}t)\Big)} -   \tanh{\Big(\Lambda(z - v_{w}t)\Big)} \Big]\,,
\end{equation}
and the interaction is assumed to be off. The collision is assumed to take place at $(z,t) = (0,0)$ and the bubbles are evolved according to the KG equation for $t>0$. The scalar is analyzed in a box of dimensions $[-L/2, L/2]$ in both space and time, with $L = 40 v_{\phi}^{-1}$. $N\times N$ samples with a base resolution $d = L/N = 0.01 v_{\phi}^{-1}$ are drawn and used to numerically solve the full KG equation to obtain $\phi(z,t)$. Subsequently, the two-dimensional discrete Fourier transform $\tilde{\phi}(\omega, k)$ is calculated, and only the real part is retained. Following Ref.~\cite{Mansour:2023fwj}, we retain only the real part of the Fourier transform, since the imaginary component is dominated by spurious finite-volume contributions near the infrared cutoff. Ref.~\cite{Mansour:2023fwj} explicitly verified that this prescription agrees closely with using the full modulus after imposing the corresponding infrared cutoff. The spatial and temporal resolutions are constrained according to
\begin{align}
	\Delta z & = \min(0.01,\, 0.2\, \Lambda^{-1},\, 0.1\, m_{\max}^{-1})\,,\\
	\Delta t & = \min(0.01,\, 0.5\, \Delta z,\, 0.1\, m_{\max}^{-1})\,,
\end{align}
where $m_{\max}(\epsilon) =\max(m_{t}(\epsilon), m_{f}(\epsilon))$. Afterward, $N_{\chi} = 180$ points are selected from the variable $\chi = \omega^{2} - k^{2}$, which defines a hyperbola in the ($\omega, k$) plane. These points are sampled from the range
\begin{equation}
	\chi \in [1, \chi_{\max}]\,, \hspace{5mm} \chi_{\max} = \min(6200, \max(150, 1.8 m_{X}^{2}))\,.
\end{equation}
Along each hyperbola, $N_{\xi}=900$ arithmetically equidistant points are selected to perform the integral in Eq.~(\ref{eq:eff_factor}) numerically and find the efficiency factor. The optimizer utilizes the least-squares method that fits the continuum first, and then tries to fit the bump. Specifically, it performs the fit in 3 sequential stages
\begin{enumerate}
\item Fit ($A, p, \Delta p, x_{c}$) outside the peak-exclusion window in log space;
\item Fit ($B,c,d$) to the positive residual in the peak window;
\item Jointly refine all 7 parameters. 
\end{enumerate}
The global bounds on the range of fit parameters are set as follows
\begin{align}
& 10^{-12} \leq A \leq 10^{10}\,, \hspace{5mm} -2 \leq p \leq 8\,, \hspace{5mm} 10^{-4} \leq \Delta p \leq 20\,, \hspace{5mm} 0.01 \leq x_{c} \leq 30\,,\nonumber\\
& \qquad \qquad 10^{-12} \leq B \leq 10^{4}\,, \hspace{5mm} -150 \leq c \leq 150\,, \hspace{5mm} 0.05 \leq d \leq 300\,.
\end{align}

We note that the seven fit parameters are correlated, and for some benchmark points one or more of them saturate the imposed fitting bounds. This does not necessarily indicate a poor fit to the numerical efficiency factor, since different parameter combinations can describe very similar spectral shapes. Indeed, all of the local benchmark fits used in the catalogue satisfy our accuracy criterion, with multiplicative RMS errors between approximately $1.14$ and $2.71$. The individual fit parameters should therefore not be assigned independent physical significance; rather, the full parameter set should be used together to reconstruct $f_X(\chi)$.

Special treatment is needed for the low-$v_{w}$ low-$\epsilon$ regime, which is difficult to fit. To improve the fit in this regime, we define the following gate functions
\begin{equation}\label{eq:gates}
g_E=S(y_v)\,S(y_\epsilon),
\qquad
g_I=0,
\end{equation}
where
\begin{equation}
y_v=\frac{0.99-v_w}{0.99-0.90},
\qquad
y_\epsilon=
\frac{\eta(0.55)-m_f/v_\phi}
{\eta(0.55)-\eta(0.40)},
\qquad
\eta(\epsilon)=\frac{m_f(\epsilon)}{v_\phi},
\end{equation}
and
\begin{equation}\label{eq:smoothstep}
S(y)=
\begin{cases}
0, & y\leq 0,\\
y^2(3-2y), & 0<y<1,\\
1, & y\geq 1.
\end{cases}
\end{equation}

The gate functions in Eqs.~(\ref{eq:gates})--(\ref{eq:smoothstep}) do not introduce an additional contribution to the efficiency factor. They are used only in the fitting procedure to smoothly interpolate the fitting windows and weights in the low-$v_w$, low-$\epsilon$ elastic region, where the numerical spectra are more difficult to fit. Thus, $g_E$ controls the strength of this adaptive fitting prescription, while $g_I=0$ because no corresponding modification is required for the inelastic fits. The spectral ansatz itself remains the continuum-plus-Gaussian form of Eq.~(\ref{eq:FitFunc}).

\newpage
\section{More results}\label{app:results}
Here we show some results for more benchmark points.
\begin{figure}[h!] 
\centering
\includegraphics[width=0.96\textwidth]{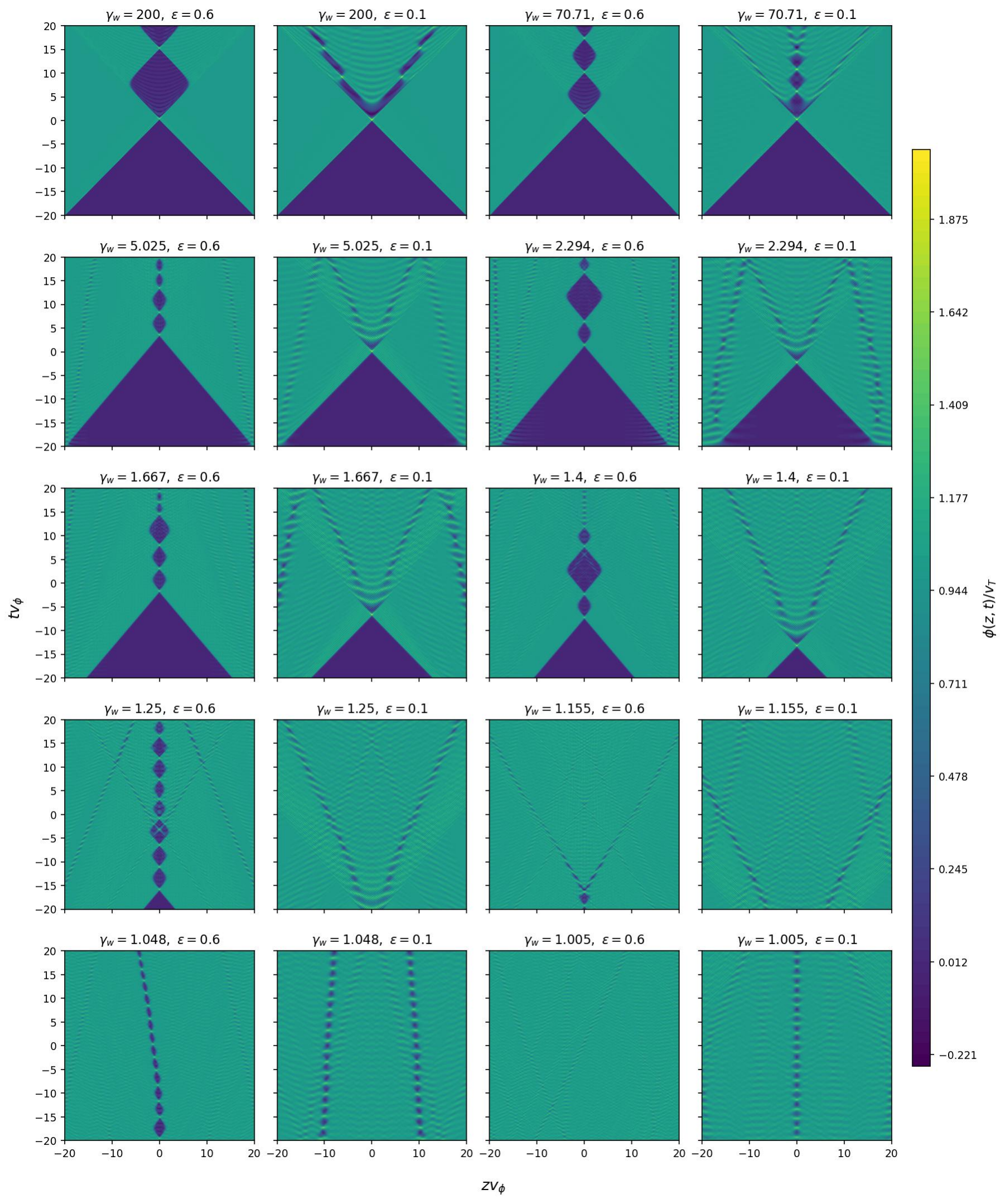}
\caption{The spacetime configuration of the scalar field for several wall velocities with $\epsilon = 0.6$ and $0.1$ corresponding to elastic and inelastic collisions, respectively. The solution corresponds to the full KG equation with a kink ansatz.}
\label{fig:MoreCollisions}
\end{figure}

\begin{figure}[h!] 
\centering
\includegraphics[width=\textwidth]{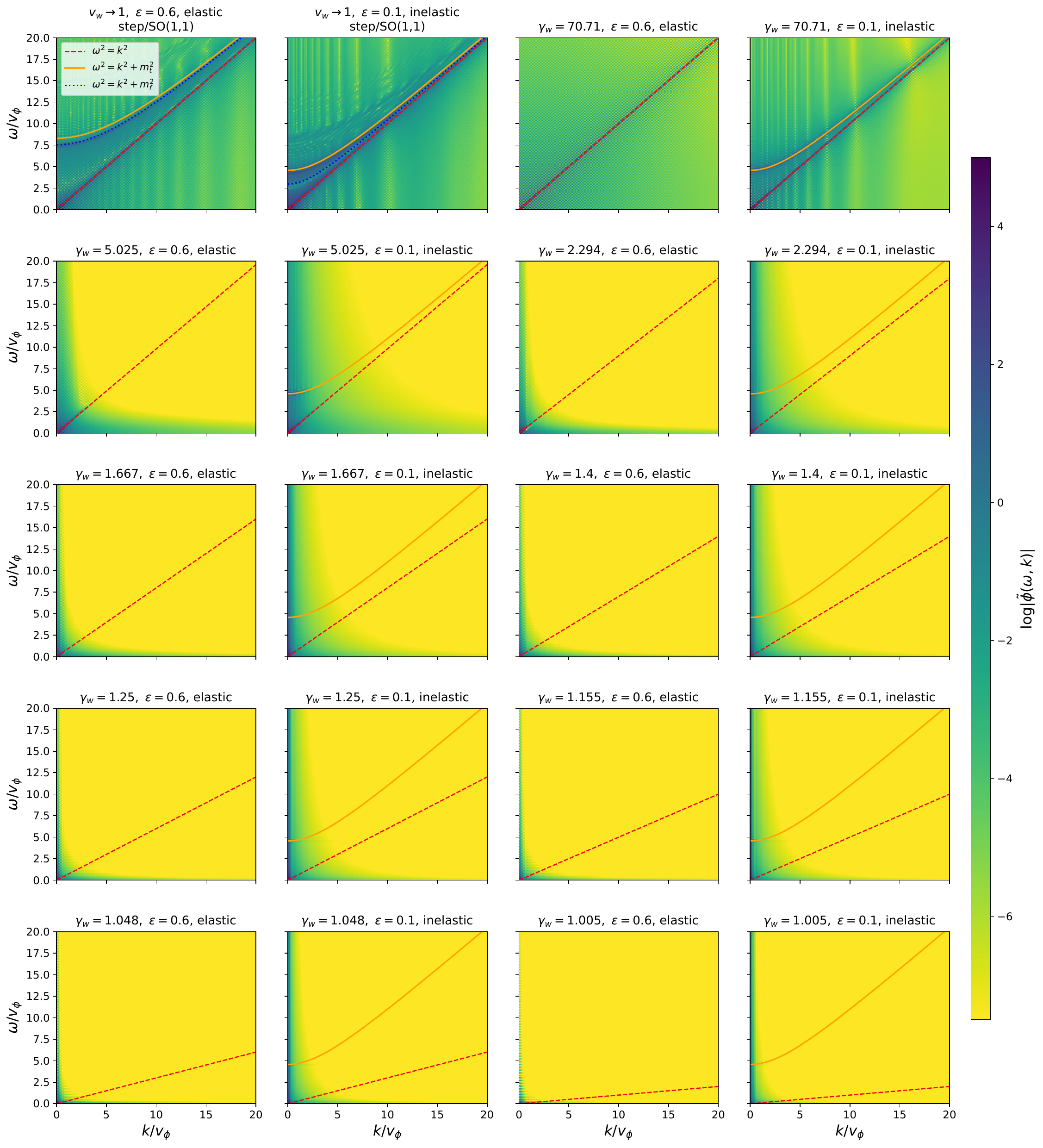}
\caption{The Fourier transform of the scalar field $|\tilde{\phi}(\omega, k)|$ for several wall velocities with $\epsilon = 0.6$ and $0.1$ corresponding to elastic and inelastic collisions, respectively. The first two plots correspond to the solution of the $SO(1,1)$ equation with a step function ansatz, whereas the remaining are obtained from the solution of the full KG equation with a kink ansatz.}
\label{fig:MoreFT}
\end{figure}

\begin{figure}[h!] 
\centering
\includegraphics[width=\textwidth]{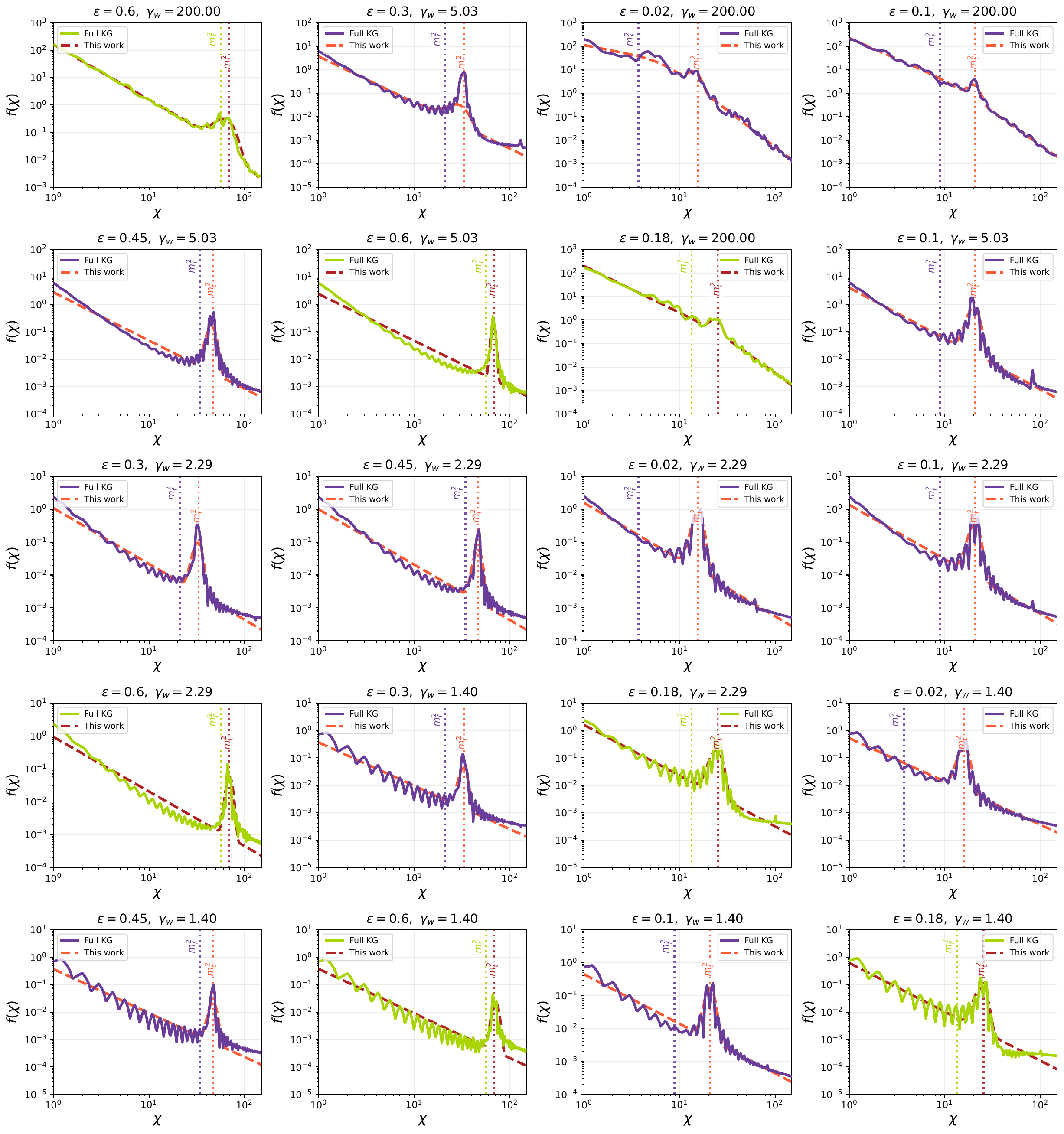}
\caption{The numerical solution of the full KG equation  with a kink ansatz, together with the fitted efficiency factors for several benchmark points in $\gamma_{w}$ and $\epsilon$.}
\label{fig:MoreFits}
\end{figure}
\newpage
\section{Fitting constant benchmark values}\label{app:constants}
Here we present the numerical values of the fitting constants $\{A_{X}, B_{X}, c_{X}, d_{X}, p_{X},\Delta p_{X},x_{c,X}\}$ for benchmark values of ($\gamma_{w}$, $\epsilon$). Here we set $v_{\phi} =1$ and $\ell_{w} = 10$ for numerical simulations. The borderline value $\epsilon=0.22$ is not used to interpolate between the elastic and inelastic branches. 

\begin{table}[h!]
\centering
\caption{Numerical-fit constants for $\gamma_w=200.00$ and $v_w=1.00$.}
\label{tab:fit-constants-gamma-200p00}
\begingroup
\scriptsize
\setlength{\tabcolsep}{2pt}
\renewcommand{\arraystretch}{0.95}
\begin{tabular*}{\textwidth}{@{\extracolsep{\fill}}rrrrrrrrrrrr@{}}
\toprule
$\epsilon$ & $v_w$ & $\gamma_w$ & $m_t$ & $m_f$ & $A$ & $p$ & $\Delta p$ & $x_c$ & $B$ & $c$ & $d$ \\
\midrule
$0.02$ & $1.00$ & $200.00$ & $3.96$ & $1.93$ & $3143.86$ & $0.79$ & $2.33$ & $0.42$ & $0.06$ & $15.56$ & $19.68$ \\
$0.04$ & $1.00$ & $200.00$ & $4.15$ & $2.29$ & $1809.05$ & $1.12$ & $1.71$ & $0.29$ & $0.06$ & $18.15$ & $22.12$ \\
$0.06$ & $1.00$ & $200.00$ & $4.30$ & $2.55$ & $705.65$ & $1.54$ & $1.31$ & $0.47$ & $0.04$ & $6.33$ & $20.52$ \\
$0.08$ & $1.00$ & $200.00$ & $4.44$ & $2.77$ & $566.32$ & $1.64$ & $1.20$ & $0.37$ & $0.03$ & $20.51$ & $23.59$ \\
$0.10$ & $1.00$ & $200.00$ & $4.56$ & $2.97$ & $449.25$ & $1.77$ & $1.06$ & $0.52$ & $0.02$ & $5.60$ & $22.91$ \\
$0.12$ & $1.00$ & $200.00$ & $4.69$ & $3.16$ & $453.23$ & $1.66$ & $0.97$ & $0.31$ & $0.02$ & $22.08$ & $33.46$ \\
$0.14$ & $1.00$ & $200.00$ & $4.80$ & $3.33$ & $444.63$ & $1.68$ & $1.09$ & $0.28$ & $0.02$ & $10.79$ & $25.95$ \\
\bottomrule
\end{tabular*}
\par\vspace{0.4em}
\endgroup
\end{table}
\clearpage

\begin{table}[h!]
\centering
\caption{Numerical-fit constants for $\gamma_w=5.03$ and $v_w=0.98$.}
\label{tab:fit-constants-gamma-5p03}
\begingroup
\scriptsize
\setlength{\tabcolsep}{2pt}
\renewcommand{\arraystretch}{0.95}
\begin{tabular*}{\textwidth}{@{\extracolsep{\fill}}rrrrrrrrrrrr@{}}
\toprule
$\epsilon$ & $v_w$ & $\gamma_w$ & $m_t$ & $m_f$ & $A$ & $p$ & $\Delta p$ & $x_c$ & $B$ & $c$ & $d$ \\
\midrule
$0.02$ & $0.98$ & $5.03$ & $3.96$ & $1.93$ & $6.48$ & $1.87$ & $1.00\!\times\!10^{-4}$ & $0.01$ & $0.01$ & $4.86$ & $18.62$ \\
$0.04$ & $0.98$ & $5.03$ & $4.15$ & $2.29$ & $4.54$ & $2.05$ & $1.04\!\times\!10^{-4}$ & $0.01$ & $0.01$ & $7.44$ & $17.69$ \\
$0.06$ & $0.98$ & $5.03$ & $4.30$ & $2.55$ & $6.37$ & $1.87$ & $1.00\!\times\!10^{-4}$ & $30.00$ & $0.01$ & $6.06$ & $18.68$ \\
$0.08$ & $0.98$ & $5.03$ & $4.44$ & $2.77$ & $4.83$ & $2.01$ & $1.00\!\times\!10^{-4}$ & $0.01$ & $0.01$ & $8.39$ & $20.16$ \\
$0.10$ & $0.98$ & $5.03$ & $4.56$ & $2.97$ & $6.29$ & $1.86$ & $1.00\!\times\!10^{-4}$ & $30.00$ & $9.38\!\times\!10^{-3}$ & $7.00$ & $20.51$ \\
$0.12$ & $0.98$ & $5.03$ & $4.69$ & $3.16$ & $6.57$ & $1.90$ & $1.00\!\times\!10^{-4}$ & $29.99$ & $0.02$ & $9.20$ & $13.45$ \\
$0.14$ & $0.98$ & $5.03$ & $4.80$ & $3.33$ & $4.88$ & $1.98$ & $1.00\!\times\!10^{-4}$ & $30.00$ & $0.01$ & $9.30$ & $17.56$ \\
\bottomrule
\end{tabular*}
\par\vspace{0.4em}
\endgroup
\end{table}
\clearpage

\begin{table}[h!]
\centering
\caption{Numerical-fit constants for $\gamma_w=2.29$ and $v_w=0.90$.}
\label{tab:fit-constants-gamma-2p29}
\begingroup
\scriptsize
\setlength{\tabcolsep}{2pt}
\renewcommand{\arraystretch}{0.95}
\begin{tabular*}{\textwidth}{@{\extracolsep{\fill}}rrrrrrrrrrrr@{}}
\toprule
$\epsilon$ & $v_w$ & $\gamma_w$ & $m_t$ & $m_f$ & $A$ & $p$ & $\Delta p$ & $x_c$ & $B$ & $c$ & $d$ \\
\midrule
$0.02$ & $0.90$ & $2.29$ & $3.96$ & $1.93$ & $3.34$ & $1.72$ & $1.00\!\times\!10^{-4}$ & $30.00$ & $6.17\!\times\!10^{-3}$ & $0.98$ & $17.81$ \\
$0.04$ & $0.90$ & $2.29$ & $4.15$ & $2.29$ & $2.81$ & $1.85$ & $1.00\!\times\!10^{-4}$ & $0.01$ & $6.36\!\times\!10^{-3}$ & $3.91$ & $15.27$ \\
$0.06$ & $0.90$ & $2.29$ & $4.30$ & $2.55$ & $3.55$ & $1.74$ & $1.00\!\times\!10^{-4}$ & $30.00$ & $6.35\!\times\!10^{-3}$ & $1.37$ & $15.46$ \\
$0.08$ & $0.90$ & $2.29$ & $4.44$ & $2.77$ & $2.53$ & $1.87$ & $1.08\!\times\!10^{-4}$ & $0.01$ & $4.98\!\times\!10^{-3}$ & $5.46$ & $18.73$ \\
$0.10$ & $0.90$ & $2.29$ & $4.56$ & $2.97$ & $3.85$ & $1.65$ & $1.00\!\times\!10^{-4}$ & $29.90$ & $4.34\!\times\!10^{-3}$ & $2.39$ & $20.35$ \\
$0.12$ & $0.90$ & $2.29$ & $4.69$ & $3.16$ & $2.62$ & $1.85$ & $1.00\!\times\!10^{-4}$ & $0.01$ & $3.67\!\times\!10^{-3}$ & $6.23$ & $20.46$ \\
$0.14$ & $0.90$ & $2.29$ & $4.80$ & $3.33$ & $4.29$ & $1.68$ & $1.00\!\times\!10^{-4}$ & $30.00$ & $6.77\!\times\!10^{-3}$ & $4.02$ & $12.62$ \\
\bottomrule
\end{tabular*}
\par\vspace{0.4em}
\endgroup
\end{table}
\clearpage

\begin{table}[h!]
\centering
\caption{Numerical-fit constants for $\gamma_w=1.67$ and $v_w=0.80$.}
\label{tab:fit-constants-gamma-1p67}
\begingroup
\scriptsize
\setlength{\tabcolsep}{2pt}
\renewcommand{\arraystretch}{0.95}
\begin{tabular*}{\textwidth}{@{\extracolsep{\fill}}rrrrrrrrrrrr@{}}
\toprule
$\epsilon$ & $v_w$ & $\gamma_w$ & $m_t$ & $m_f$ & $A$ & $p$ & $\Delta p$ & $x_c$ & $B$ & $c$ & $d$ \\
\midrule
$0.02$ & $0.80$ & $1.67$ & $3.96$ & $1.93$ & $2.31$ & $1.64$ & $1.00\!\times\!10^{-4}$ & $29.92$ & $3.59\!\times\!10^{-3}$ & $-0.27$ & $17.48$ \\
$0.04$ & $0.80$ & $1.67$ & $4.15$ & $2.29$ & $2.09$ & $1.75$ & $1.00\!\times\!10^{-4}$ & $0.01$ & $3.78\!\times\!10^{-3}$ & $3.96$ & $14.26$ \\
$0.06$ & $0.80$ & $1.67$ & $4.30$ & $2.55$ & $2.50$ & $1.67$ & $1.00\!\times\!10^{-4}$ & $30.00$ & $4.20\!\times\!10^{-3}$ & $-0.43$ & $14.25$ \\
$0.08$ & $0.80$ & $1.67$ & $4.44$ & $2.77$ & $1.74$ & $1.81$ & $1.17\!\times\!10^{-4}$ & $0.01$ & $2.81\!\times\!10^{-3}$ & $5.94$ & $18.44$ \\
$0.10$ & $0.80$ & $1.67$ & $4.56$ & $2.97$ & $2.74$ & $1.57$ & $1.00\!\times\!10^{-4}$ & $30.00$ & $2.71\!\times\!10^{-3}$ & $0.16$ & $19.29$ \\
$0.12$ & $0.80$ & $1.67$ & $4.69$ & $3.16$ & $1.78$ & $1.80$ & $1.00\!\times\!10^{-4}$ & $0.01$ & $2.20\!\times\!10^{-3}$ & $7.01$ & $20.40$ \\
$0.14$ & $0.80$ & $1.67$ & $4.80$ & $3.33$ & $3.00$ & $1.60$ & $1.00\!\times\!10^{-4}$ & $30.00$ & $3.67\!\times\!10^{-3}$ & $3.16$ & $12.83$ \\
\bottomrule
\end{tabular*}
\par\vspace{0.4em}
\endgroup
\end{table}
\clearpage

\begin{table}[h!]
\centering
\caption{Numerical-fit constants for $\gamma_w=1.40$ and $v_w=0.70$.}
\label{tab:fit-constants-gamma-1p40}
\begingroup
\scriptsize
\setlength{\tabcolsep}{2pt}
\renewcommand{\arraystretch}{0.95}
\begin{tabular*}{\textwidth}{@{\extracolsep{\fill}}rrrrrrrrrrrr@{}}
\toprule
$\epsilon$ & $v_w$ & $\gamma_w$ & $m_t$ & $m_f$ & $A$ & $p$ & $\Delta p$ & $x_c$ & $B$ & $c$ & $d$ \\
\midrule
$0.02$ & $0.70$ & $1.40$ & $3.96$ & $1.93$ & $1.67$ & $1.58$ & $1.00\!\times\!10^{-4}$ & $30.00$ & $2.37\!\times\!10^{-3}$ & $-1.27$ & $17.20$ \\
$0.04$ & $0.70$ & $1.40$ & $4.15$ & $2.29$ & $1.51$ & $1.70$ & $1.04\!\times\!10^{-4}$ & $0.01$ & $2.45\!\times\!10^{-3}$ & $4.33$ & $13.90$ \\
$0.06$ & $0.70$ & $1.40$ & $4.30$ & $2.55$ & $1.83$ & $1.61$ & $1.00\!\times\!10^{-4}$ & $29.95$ & $3.11\!\times\!10^{-3}$ & $-1.12$ & $13.06$ \\
$0.08$ & $0.70$ & $1.40$ & $4.44$ & $2.77$ & $1.27$ & $1.77$ & $4.62\!\times\!10^{-4}$ & $0.01$ & $1.79\!\times\!10^{-3}$ & $6.74$ & $18.26$ \\
$0.10$ & $0.70$ & $1.40$ & $4.56$ & $2.97$ & $2.00$ & $1.51$ & $1.00\!\times\!10^{-4}$ & $30.00$ & $1.93\!\times\!10^{-3}$ & $-1.52$ & $18.32$ \\
$0.12$ & $0.70$ & $1.40$ & $4.69$ & $3.16$ & $1.29$ & $1.77$ & $1.00\!\times\!10^{-4}$ & $0.01$ & $1.48\!\times\!10^{-3}$ & $8.01$ & $20.12$ \\
$0.14$ & $0.70$ & $1.40$ & $4.80$ & $3.33$ & $2.13$ & $1.55$ & $1.00\!\times\!10^{-4}$ & $30.00$ & $2.30\!\times\!10^{-3}$ & $2.57$ & $12.81$ \\
\bottomrule
\end{tabular*}
\par\vspace{0.4em}
\endgroup
\end{table}
\clearpage

\begin{table}[h!]
\centering
\caption{Numerical-fit constants for $\gamma_w=1.25$ and $v_w=0.60$.}
\label{tab:fit-constants-gamma-1p25}
\begingroup
\scriptsize
\setlength{\tabcolsep}{2pt}
\renewcommand{\arraystretch}{0.95}
\begin{tabular*}{\textwidth}{@{\extracolsep{\fill}}rrrrrrrrrrrr@{}}
\toprule
$\epsilon$ & $v_w$ & $\gamma_w$ & $m_t$ & $m_f$ & $A$ & $p$ & $\Delta p$ & $x_c$ & $B$ & $c$ & $d$ \\
\midrule
$0.02$ & $0.60$ & $1.25$ & $3.96$ & $1.93$ & $1.22$ & $1.53$ & $1.00\!\times\!10^{-4}$ & $30.00$ & $1.67\!\times\!10^{-3}$ & $-2.57$ & $16.53$ \\
$0.04$ & $0.60$ & $1.25$ & $4.15$ & $2.29$ & $1.08$ & $1.66$ & $1.00\!\times\!10^{-4}$ & $0.01$ & $1.59\!\times\!10^{-3}$ & $4.68$ & $13.71$ \\
$0.06$ & $0.60$ & $1.25$ & $4.30$ & $2.55$ & $1.32$ & $1.56$ & $1.00\!\times\!10^{-4}$ & $29.95$ & $2.15\!\times\!10^{-3}$ & $-1.08$ & $12.63$ \\
$0.08$ & $0.60$ & $1.25$ & $4.44$ & $2.77$ & $0.92$ & $1.74$ & $1.01\!\times\!10^{-4}$ & $0.01$ & $1.18\!\times\!10^{-3}$ & $7.11$ & $18.30$ \\
$0.10$ & $0.60$ & $1.25$ & $4.56$ & $2.97$ & $1.39$ & $1.46$ & $1.00\!\times\!10^{-4}$ & $30.00$ & $1.23\!\times\!10^{-3}$ & $-2.18$ & $18.19$ \\
$0.12$ & $0.60$ & $1.25$ & $4.69$ & $3.16$ & $0.94$ & $1.74$ & $6.19\!\times\!10^{-4}$ & $0.01$ & $1.01\!\times\!10^{-3}$ & $8.94$ & $19.75$ \\
$0.14$ & $0.60$ & $1.25$ & $4.80$ & $3.33$ & $1.52$ & $1.51$ & $1.00\!\times\!10^{-4}$ & $30.00$ & $1.49\!\times\!10^{-3}$ & $2.09$ & $12.71$ \\
\bottomrule
\end{tabular*}
\par\vspace{0.4em}
\endgroup
\end{table}
\clearpage

\begin{table}[h!]
\centering
\caption{Numerical-fit constants for $\gamma_w=1.15$ and $v_w=0.50$.}
\label{tab:fit-constants-gamma-1p15}
\begingroup
\scriptsize
\setlength{\tabcolsep}{2pt}
\renewcommand{\arraystretch}{0.95}
\begin{tabular*}{\textwidth}{@{\extracolsep{\fill}}rrrrrrrrrrrr@{}}
\toprule
$\epsilon$ & $v_w$ & $\gamma_w$ & $m_t$ & $m_f$ & $A$ & $p$ & $\Delta p$ & $x_c$ & $B$ & $c$ & $d$ \\
\midrule
$0.02$ & $0.50$ & $1.15$ & $3.96$ & $1.93$ & $0.84$ & $1.49$ & $1.00\!\times\!10^{-4}$ & $29.95$ & $1.09\!\times\!10^{-3}$ & $-3.01$ & $16.42$ \\
$0.04$ & $0.50$ & $1.15$ & $4.15$ & $2.29$ & $0.74$ & $1.63$ & $5.94\!\times\!10^{-4}$ & $0.01$ & $1.01\!\times\!10^{-3}$ & $5.65$ & $13.97$ \\
$0.06$ & $0.50$ & $1.15$ & $4.30$ & $2.55$ & $0.90$ & $1.54$ & $1.00\!\times\!10^{-4}$ & $29.98$ & $1.41\!\times\!10^{-3}$ & $-1.15$ & $12.59$ \\
$0.08$ & $0.50$ & $1.15$ & $4.44$ & $2.77$ & $0.65$ & $1.71$ & $3.10\!\times\!10^{-3}$ & $0.01$ & $7.52\!\times\!10^{-4}$ & $7.70$ & $18.20$ \\
$0.10$ & $0.50$ & $1.15$ & $4.56$ & $2.97$ & $0.92$ & $1.42$ & $1.00\!\times\!10^{-4}$ & $30.00$ & $8.32\!\times\!10^{-4}$ & $-2.91$ & $17.84$ \\
$0.12$ & $0.50$ & $1.15$ & $4.69$ & $3.16$ & $0.66$ & $1.72$ & $3.76\!\times\!10^{-4}$ & $0.01$ & $6.78\!\times\!10^{-4}$ & $9.71$ & $19.36$ \\
$0.14$ & $0.50$ & $1.15$ & $4.80$ & $3.33$ & $1.01$ & $1.48$ & $1.00\!\times\!10^{-4}$ & $29.95$ & $8.91\!\times\!10^{-4}$ & $3.35$ & $13.91$ \\
\bottomrule
\end{tabular*}
\par\vspace{0.4em}
\endgroup
\end{table}
\clearpage

\begin{table}[h!]
\centering
\caption{Numerical-fit constants for $\gamma_w=1.09$ and $v_w=0.40$.}
\label{tab:fit-constants-gamma-1p09}
\begingroup
\scriptsize
\setlength{\tabcolsep}{2pt}
\renewcommand{\arraystretch}{0.95}
\begin{tabular*}{\textwidth}{@{\extracolsep{\fill}}rrrrrrrrrrrr@{}}
\toprule
$\epsilon$ & $v_w$ & $\gamma_w$ & $m_t$ & $m_f$ & $A$ & $p$ & $\Delta p$ & $x_c$ & $B$ & $c$ & $d$ \\
\midrule
$0.02$ & $0.40$ & $1.09$ & $3.96$ & $1.93$ & $0.57$ & $1.47$ & $1.00\!\times\!10^{-4}$ & $29.95$ & $7.30\!\times\!10^{-4}$ & $-4.30$ & $15.47$ \\
$0.04$ & $0.40$ & $1.09$ & $4.15$ & $2.29$ & $0.49$ & $1.61$ & $6.39\!\times\!10^{-4}$ & $0.01$ & $6.04\!\times\!10^{-4}$ & $5.70$ & $14.00$ \\
$0.06$ & $0.40$ & $1.09$ & $4.30$ & $2.55$ & $0.57$ & $1.52$ & $1.00\!\times\!10^{-4}$ & $30.00$ & $8.65\!\times\!10^{-4}$ & $-1.19$ & $12.52$ \\
$0.08$ & $0.40$ & $1.09$ & $4.44$ & $2.77$ & $0.42$ & $1.70$ & $1.00\!\times\!10^{-4}$ & $0.01$ & $4.54\!\times\!10^{-4}$ & $8.14$ & $18.13$ \\
$0.10$ & $0.40$ & $1.09$ & $4.56$ & $2.97$ & $0.57$ & $1.39$ & $1.00\!\times\!10^{-4}$ & $30.00$ & $5.10\!\times\!10^{-4}$ & $-3.48$ & $17.69$ \\
$0.12$ & $0.40$ & $1.09$ & $4.69$ & $3.16$ & $0.43$ & $1.71$ & $2.80\!\times\!10^{-4}$ & $0.01$ & $4.25\!\times\!10^{-4}$ & $10.27$ & $19.01$ \\
$0.14$ & $0.40$ & $1.09$ & $4.80$ & $3.33$ & $0.66$ & $1.47$ & $1.00\!\times\!10^{-4}$ & $30.00$ & $5.30\!\times\!10^{-4}$ & $3.17$ & $13.95$ \\
\bottomrule
\end{tabular*}
\par\vspace{0.4em}
\endgroup
\end{table}
\clearpage

\begin{table}[h!]
\centering
\caption{Numerical-fit constants for $\gamma_w=1.05$ and $v_w=0.30$.}
\label{tab:fit-constants-gamma-1p05}
\begingroup
\scriptsize
\setlength{\tabcolsep}{2pt}
\renewcommand{\arraystretch}{0.95}
\begin{tabular*}{\textwidth}{@{\extracolsep{\fill}}rrrrrrrrrrrr@{}}
\toprule
$\epsilon$ & $v_w$ & $\gamma_w$ & $m_t$ & $m_f$ & $A$ & $p$ & $\Delta p$ & $x_c$ & $B$ & $c$ & $d$ \\
\midrule
$0.02$ & $0.30$ & $1.05$ & $3.96$ & $1.93$ & $0.32$ & $1.46$ & $1.00\!\times\!10^{-4}$ & $30.00$ & $4.03\!\times\!10^{-4}$ & $-4.65$ & $15.30$ \\
$0.04$ & $0.30$ & $1.05$ & $4.15$ & $2.29$ & $0.28$ & $1.60$ & $1.04\!\times\!10^{-4}$ & $0.01$ & $3.22\!\times\!10^{-4}$ & $5.71$ & $14.04$ \\
$0.06$ & $0.30$ & $1.05$ & $4.30$ & $2.55$ & $0.32$ & $1.51$ & $1.00\!\times\!10^{-4}$ & $29.96$ & $4.73\!\times\!10^{-4}$ & $-1.22$ & $12.48$ \\
$0.08$ & $0.30$ & $1.05$ & $4.44$ & $2.77$ & $0.25$ & $1.69$ & $1.08\!\times\!10^{-4}$ & $0.01$ & $2.45\!\times\!10^{-4}$ & $8.43$ & $18.10$ \\
$0.10$ & $0.30$ & $1.05$ & $4.56$ & $2.97$ & $0.32$ & $1.38$ & $1.00\!\times\!10^{-4}$ & $30.00$ & $2.78\!\times\!10^{-4}$ & $-3.92$ & $17.57$ \\
$0.12$ & $0.30$ & $1.05$ & $4.69$ & $3.16$ & $0.25$ & $1.71$ & $1.03\!\times\!10^{-4}$ & $0.01$ & $2.37\!\times\!10^{-4}$ & $10.65$ & $18.74$ \\
$0.14$ & $0.30$ & $1.05$ & $4.80$ & $3.33$ & $0.37$ & $1.47$ & $1.00\!\times\!10^{-4}$ & $30.00$ & $2.39\!\times\!10^{-4}$ & $1.78$ & $15.52$ \\
\bottomrule
\end{tabular*}
\par\vspace{0.4em}
\endgroup
\end{table}
\clearpage

\begin{table}[h!]
\centering
\caption{Numerical-fit constants for $\gamma_w=1.02$ and $v_w=0.20$.}
\label{tab:fit-constants-gamma-1p02}
\begingroup
\scriptsize
\setlength{\tabcolsep}{2pt}
\renewcommand{\arraystretch}{0.95}
\begin{tabular*}{\textwidth}{@{\extracolsep{\fill}}rrrrrrrrrrrr@{}}
\toprule
$\epsilon$ & $v_w$ & $\gamma_w$ & $m_t$ & $m_f$ & $A$ & $p$ & $\Delta p$ & $x_c$ & $B$ & $c$ & $d$ \\
\midrule
$0.02$ & $0.20$ & $1.02$ & $3.96$ & $1.93$ & $0.15$ & $1.46$ & $1.00\!\times\!10^{-4}$ & $30.00$ & $1.77\!\times\!10^{-4}$ & $-4.90$ & $15.17$ \\
$0.04$ & $0.20$ & $1.02$ & $4.15$ & $2.29$ & $0.13$ & $1.60$ & $1.01\!\times\!10^{-4}$ & $0.01$ & $1.38\!\times\!10^{-4}$ & $5.71$ & $14.07$ \\
$0.06$ & $0.20$ & $1.02$ & $4.30$ & $2.55$ & $0.15$ & $1.50$ & $1.00\!\times\!10^{-4}$ & $30.00$ & $2.06\!\times\!10^{-4}$ & $-1.26$ & $12.45$ \\
$0.08$ & $0.20$ & $1.02$ & $4.44$ & $2.77$ & $0.11$ & $1.69$ & $1.01\!\times\!10^{-4}$ & $0.01$ & $1.06\!\times\!10^{-4}$ & $8.60$ & $18.10$ \\
$0.10$ & $0.20$ & $1.02$ & $4.56$ & $2.97$ & $0.14$ & $1.38$ & $1.00\!\times\!10^{-4}$ & $29.92$ & $1.21\!\times\!10^{-4}$ & $-4.23$ & $17.46$ \\
$0.12$ & $0.20$ & $1.02$ & $4.69$ & $3.16$ & $0.12$ & $1.71$ & $1.11\!\times\!10^{-3}$ & $0.01$ & $1.05\!\times\!10^{-4}$ & $10.89$ & $18.54$ \\
$0.14$ & $0.20$ & $1.02$ & $4.80$ & $3.33$ & $0.17$ & $1.46$ & $1.00\!\times\!10^{-4}$ & $30.00$ & $1.03\!\times\!10^{-4}$ & $1.66$ & $15.54$ \\
\bottomrule
\end{tabular*}
\par\vspace{0.4em}
\endgroup
\end{table}
\clearpage

\begin{table}[h!]
\centering
\caption{Numerical-fit constants for $\gamma_w=1.01$ and $v_w=0.10$.}
\label{tab:fit-constants-gamma-1p01}
\begingroup
\scriptsize
\setlength{\tabcolsep}{2pt}
\renewcommand{\arraystretch}{0.95}
\begin{tabular*}{\textwidth}{@{\extracolsep{\fill}}rrrrrrrrrrrr@{}}
\toprule
$\epsilon$ & $v_w$ & $\gamma_w$ & $m_t$ & $m_f$ & $A$ & $p$ & $\Delta p$ & $x_c$ & $B$ & $c$ & $d$ \\
\midrule
$0.02$ & $0.10$ & $1.01$ & $3.96$ & $1.93$ & $0.04$ & $1.46$ & $1.00\!\times\!10^{-4}$ & $30.00$ & $4.39\!\times\!10^{-5}$ & $-5.05$ & $15.10$ \\
$0.04$ & $0.10$ & $1.01$ & $4.15$ & $2.29$ & $0.03$ & $1.60$ & $1.02\!\times\!10^{-4}$ & $0.02$ & $3.37\!\times\!10^{-5}$ & $5.71$ & $14.10$ \\
$0.06$ & $0.10$ & $1.01$ & $4.30$ & $2.55$ & $0.04$ & $1.50$ & $1.00\!\times\!10^{-4}$ & $29.90$ & $5.10\!\times\!10^{-5}$ & $-1.31$ & $12.44$ \\
$0.08$ & $0.10$ & $1.01$ & $4.44$ & $2.77$ & $0.03$ & $1.69$ & $1.90\!\times\!10^{-3}$ & $0.01$ & $2.61\!\times\!10^{-5}$ & $8.69$ & $18.11$ \\
$0.10$ & $0.10$ & $1.01$ & $4.56$ & $2.97$ & $0.04$ & $1.38$ & $1.00\!\times\!10^{-4}$ & $30.00$ & $2.99\!\times\!10^{-5}$ & $-4.43$ & $17.38$ \\
$0.12$ & $0.10$ & $1.01$ & $4.69$ & $3.16$ & $0.03$ & $1.71$ & $1.01\!\times\!10^{-4}$ & $0.01$ & $2.62\!\times\!10^{-5}$ & $11.02$ & $18.41$ \\
$0.14$ & $0.10$ & $1.01$ & $4.80$ & $3.33$ & $0.04$ & $1.47$ & $1.00\!\times\!10^{-4}$ & $30.00$ & $2.53\!\times\!10^{-5}$ & $1.59$ & $15.53$ \\
\bottomrule
\end{tabular*}
\par\vspace{0.4em}
\endgroup
\end{table}

\end{document}